\documentclass[useAMS,usenatbib]{mn2e}
\usepackage{graphicx,amsmath,multirow,amssymb}
\usepackage{natbib}
\usepackage{subfigure}
\newcommand{\comment}[1]{}

\def\simgt{\lower.5ex\hbox{$\; \buildrel > \over \sim \;$}}
\def\simlt{\lower.5ex\hbox{$\; \buildrel < \over \sim \;$}}

\title[Alumina dust in the winds of O-rich AGBs]{On the alumina dust production in the winds of 
O-rich Asymptotic Giant Branch stars}
\author[Dell'Agli et al.]{F. Dell'Agli$^{1,2}$, D. A. Garc\'{\i}a-Hern\'andez$^{3,4}$, C.
Rossi$^{2}$, P. Ventura$^1$, 
\newauthor
M. Di Criscienzo$^1$, R. Schneider$^1$\\
$^1$INAF -- Osservatorio Astronomico di Roma, Via Frascati 33, 00040, Monte Porzio Catone (RM), Italy \\
$^{2}$Dipartimento di Fisica, Universit\`a di Roma ``La Sapienza'', P.le Aldo Moro 5, 00143, Roma, Italy \\
$^{3}$Instituto de Astrof\'{\i}sica de Canarias, V\'{\i}a L\'actea s/n, E-38200 La Laguna, Tenerife, Spain \\
$^{4}$Departamento de Astrof\'{\i}sica, Universidad de La Laguna (ULL), E-38206 La Laguna, Spain}

\begin{document}

\date{Accepted 2013 xx xx., Received 2013 xx xx; in original form 2013 xx xx}

\pagerange{\pageref{firstpage}--\pageref{lastpage}} \pubyear{2013}

\maketitle

\label{firstpage}

\begin{abstract}
The O-rich Asymptotic Giant Branch (AGB) stars experience strong mass 
loss with efficient dust condensation and they are major sources 
of dust in the interstellar medium. Alumina dust (Al$_2$O$_3$) is an 
important dust component in O-rich circumstellar shells and it is 
expected to be fairly abundant in the winds of the more massive and 
O-rich AGB
stars. By coupling AGB 
stellar nucleosynthesis and dust formation, we present a self-consistent exploration on 
the Al$_2$O$_3$ production in 
the winds of AGB stars with progenitor masses between $\sim$3 and 7 M$_{\odot}$ 
and metallicities in the range 0.0003 $\le$ Z $\le$ 0.018. We 
find that Al$_2$O$_3$ particles form at radial distances from the
centre between $\sim2$ and 4 R$_*$ (depending on metallicity), which is in 
agreement with recent interferometric observations of Galactic O-rich 
AGB stars. The mass of Al$_2$O$_3$ dust is found to scale almost linearly 
with metallicity, with solar metallicity AGBs producing the 
highest amount (about 10$^{-3}$ M$_{\odot}$) of alumina dust. The Al$_2$O$_3$ grain size 
decreases with decreasing metallicity (and initial stellar mass) and the maximum size of the Al$_2$O$_3$ grains 
is $\sim$0.075 $\mu m$ for the solar metallicity models. Interestingly, the strong depletion of gaseous Al 
observed in the low-metallicity HBB AGB star HV 2576 seems to be 
consistent with the formation of Al$_2$O$_3$ dust as predicted by our 
models. We suggest that the content of Al may be used as a mass (and 
evolutionary stage) indicator in AGB stars experiencing HBB.

\end{abstract}


\section{Introduction}

During the Asymptotic Giant Branch (AGB) phase, stars of low-- and intermediate--mass
($1M_{\odot} \le M \le 8 M_{\odot}$) experience high mass-loss
rates \citep{herwig05}, thus an efficient dust condensation \citep{fg01, fg02, fg06}
occurs in their circumstellar envelopes. In fact, they are one of
the most important contributors of dust to the interstellar medium. We note that
core-collapse supernovae (SNe) are also major producers of dust grains
\citep{matsu11,gomez12,ten-tag13}; indeed the study of the stellar source/s (e.g., AGB
stars and/or SNe) for interstellar and presolar dust grains is a hot topic of
extreme interest for the astrophysical community (Trigo-Rodr\'{\i}guez
et al. 2009, Valiante et al. 2009, Gall et al. 2011).

AGB stars experience a series of periodic, thermally unstable, ignitions of
the He shell, in what is commonly known as thermal pulse (TP) \citep{schw65, schw67}. 
The third dredge-up (TDU), occurring during the TP-AGB phase, may alter the surface chemical
abundances in AGBs, favoring a gradual carbon enrichment of the external
layers, until the originally O-rich star becomes C-rich (C/O $>$ 1). The more
massive ($> 3-4$ M$_{\odot}$) stars, however, may remain O-rich during the whole AGB
evolution as a consequence of the activation of the hot bottom burning (HBB)
process (see e.g., Bl\"ocker \& Sch\"onberner 1991, Mazzitelli et al. 1999; 
Garc\'{\i}a-Hern\'andez et al. 2007a and references therein). Thus, the dominant 
chemistry of dust grains
produced by AGB stars mainly depends on their progenitor masses and
metallicities \citep{paperI, paperII, paperIV}. Low mass AGBs (M $\le 3
M_{\odot}$) produce oxygen-rich dust as far as the surface C/O is below unity. However, this 
dust production is in limited quantities, given the low-mass loss rates experienced 
during these evolutionary phases. These stars are expected to form considerable 
amount of carbonaceous dust once the C/O ratio exceeds unity, because of the large 
number of carbon molecules available and the higher mass-loss rate experienced towards 
the end of the AGB phase \citep{wachter08}. 
On the other hand, more massive AGBs - where HBB prevents the formation of carbon stars -
produce oxygen-based dust (such as amorphous and crystalline silicates,
Sylvester et al. 1999; Garc\'{\i}a-Hern\'andez et al. 2007b). 

The dust condensation sequence as well as the most important molecules in the
nucleation process are different for C- and O--rich AGB stars. The formation of
dust grains in the winds of C-rich AGBs seems to be rather well established with
several forms of carbon (e.g., small hydrocarbon molecules such as acetylene)
believed to be the building blocks of more complex organic molecules and grains
(Cherchneff \& Cau 1999). However, the process of dust formation and grain growth in
O--rich AGB stars is less clear (e.g., Woitke 2006; Norris et al. 2012;
Zhao-Geisler et al. 2012) and needs further observational and theoretical
efforts.

The spectral energy distributions (SEDs) observed in O--rich AGB stars can be
reproduced with alumina (Al$_2$O$_3$) dust envelopes, amorphous silicates dust
shells, or a mix of both species (e.g., Lorenz-Martins \& Pompeia 2000; Maldoni
et al. 2005). In addition, the 13 $\mu$m dust emission feature generally
observed in the spectra of O-rich AGBs is attributed to corundum (crystalline
Al$_2$O$_3$) dust grains \citep{Posch99, Sloan03, DePew06, yang08, takigawa09,
Zeidler13, Jones14}. This shows that aluminium--based dust species represent a major dust
component in O--rich circumstellar shells. Al$_2$O$_3$ dust grains - because of
their high stability and transparency - have been suggested as good candidates
to explain the dusty regions observed extremely close to the stellar surface
\citep{woitke06, nor12, Zhao12}. Furthermore, Infrared Space
Observatory (ISO) spectra of Galactic bulge AGBs were found to display 
alumina dust component much stronger than silicates ones, suggesting 
Al$_2$O$_3$ grains as a likely starting point in O-rich dust condensation
sequence \citep{Blommaert06}. By following these earlier studies, Karovicova et
al. (2013) have recently analysed the mid-IR multi-epoch
interferometric observations of several O-rich AGB stars. Interestingly, their
interferometric results indicate that Al$_2$O$_3$ grains condense very close to
the stellar surface (at about 2 stellar radii which is much closer than warm
silicate grains\footnote{Note that this is supported by theoretical
thermodynamic calculations that show that Al$_2$O$_3$ condenses at higher
temperatures ($\sim$1400 K) than several types of silicates ($\sim$1100 K)
(e.g., Tielens et al. 1998)}) and that they can be seed particles for the
further O-based dust condensation.

The more massive and O-rich HBB AGB stars have been suggested as major alumina 
dust producers \citep{sedlmayr89, gs98}. The strong HBB experienced by these
stars would favour a considerable increase in the Al content at the stellar
surface because of the activation of the Mg--Al nucleosynthesis at the bottom of
the convective envelope. Presolar alumina grains originated from AGB stars were found in primitive
chondrites (Hutcheon et al. 1994, Choi et al. 1998, Nittler et al. 2008,
Takigawa et al. 2014), which is direct evidence of alumina formation around AGB
stars.  In addition, an 26Al excess is widely found in primitive refractory
materials (such as calcium and aliminium-rich inclusions; CAIs), and this 26Al excess may be
explained by pollution from a nearby massive AGB and/or super-AGB star during
the Early Solar System (ESS) (see e.g., Trigo-Rodrguez et al. 2009; Lugaro et al.
2012, and references therein).
Despite their broad astrophysical interest, a self-consistent
investigation (i.e., by coupling AGB stellar nucleosynthesis and dust formation)
on the production of Al$_2$O$_3$ in the surroundings of AGB stars is still
lacking in the literature. 

In the present work, we attempt to fill this gap, by investigating the
formation of alumina dust in the winds of AGBs. We restrict our attention to O--rich stars, 
spanning the range of metallicities $3\times 10^{-4} \leq Z \leq 0.018$.
The paper is organized as follows: the details of stellar evolution modelling
and the description of dust formation process in AGB winds are presented in
Section 2; Section 3 is focused on the properties of the Al$_2$O$_3$ molecule and on
the Mg--Al nucleosynthesis in HBB AGBs, which determines the change in the
surface content of aluminium. The formation and growth of alumina dust grains 
is described in Section 4 together with a discussion on the uncertainties of the results,
due in particular to the poor knowledge of the sticking coefficient; Section 5 presents 
a comparisons of our results with the observations available in the literature. Finally, Section 6 we draw the main conclusions of the present study.   

\section{THE MODEL}

\subsection{Stellar evolution models}

We computed the stellar evolutionary sequences for different progenitor masses
($3M_{\odot}\le M \le7M_{\odot}$) and metallicities ($3\times 10^{-4}$ $\le$ Z
$\le$ 0.018, see below) by using the code ATON. We refer the reader to
\citet{ventura98} and \citet{vd09} for a complete description of the numerical
structure of the code and of the most recent updates, respectively. We report
here only the  physical inputs relevant for the present
investigation.

The temperature gradient in regions unstable to convection was determined by the
Full Spectrum of Turbulence (FST) model presented by \citet{cm91}. This choice
is crucial for the results obtained, because of the great impact of the
convection model used on the description of the AGB evolution \citep{vd05,
dagh13}, and, consequently, on the type of dust formed around massive AGBs
\citep{paperI, paperII, paperIV}.

Mass loss was modeled according to the treatment discussed in \citet{blocker95}.
This choice is extremely relevant in order to determine the amount of dust
formed, because, as we will discuss in the following Section, the mass-loss rate
determines the density of the gas in the wind and thus the number of gas
molecules potentially able to condense into dust.

The chemical mixtures used to define the initial composition of the models were chosen
following \citet{grevesse98}. For the Z$=3\times 10^{-4}$ case we used an $\alpha-$enhancement
$[\alpha/{\rm Fe}]=+0.4$, for the metallicities Z$=4\times 10^{-3}$ and Z$=8\times 10^{-3}$
we adopted $[\alpha/{\rm Fe}]=+0.2$, whereas for Z=0.018 we used a solar--scaled mixture.

\subsection{Wind structure and dust formation}

We model the structure of the wind by following the schematisation  described in
the series of papers by \citet{fg01, fg02, fg06}. We summarize here  the main
aspects of this model.

Based on the results from stellar evolution modelling, (i.e., the temporal
evolution of mass, $M$, luminosity, $L$, effective temperature, $T_{eff}$, and mass-loss rate of the
star, $\dot{M}$), and assuming an isotropically expanding wind, we integrate a set of
equations to determine the radial distribution of velocity ($v$), temperature ($T$),
density ($\rho$), and opacity ($k$) of gas molecules. 

From the equation of momentum conservation we determine the radial velocity gradient
of the wind (in the following, we indicate with $r$ the distance from the centre of
the star): 

\begin{equation}
v{dv\over dr}=-{GM\over r^2}(1-\Gamma).
\label{eqvel}
\end{equation}

\noindent$\Gamma$ is the ratio between the radiation pressure on the dust grains and the
gravitational pull:

\begin{equation}
\Gamma={kL\over 4\pi cGM}
\label{eqgamma}
\end{equation}
\noindent

\noindent and $k$ is the flux-averaged extinction coefficient

\begin{equation}
k=k_{\rm gas}+\sum_i f_ik_i.
\label{eqk}
\end{equation}
\noindent

\noindent Eq. \ref{eqk} contains the gas contribution $k_{gas}=10^{-8}\rho^{2/3}T^3$ (Bell
\& Lin 1994), and the sum of the absorption and scattering coefficients extended
to all the dust species  considered. 

The $f_{i}$'s in Eq. \ref{eqk} are the degrees of condensation of the
key elements for each  dust species, whereas the $k_i$'s are the corresponding
extinction coefficients.

When $k$ increases, owing to dust formation, and $\Gamma$ becomes greater than
unity, the wind is accelerated by the radiation pressure. In the equation
describing the radial variation of velocity we neglect pressure forces,  because
they are negligible compared to gravity. This assumption holds in the  present
treatment, because we do not consider any shock structure of the outflow.

Starting from mass conservation, we can write the density of the wind as:

\begin{equation}
\dot M=4\pi r^2 \rho v,
\label{eqmloss}
\end{equation}
\noindent

\noindent The temperature stratification is determined assuming the grey atmosphere
approximation:

\begin{equation}
T^4={1\over 2}T_{\rm eff}^4 \left[ 1-\sqrt{1-{R_*^2\over r^2}}+{3\over 2}\tau \right],
\label{eqteff}
\end{equation}
\noindent

\noindent where $R_{*}$ is the stellar radius. The optical depth, $\tau$, is found via the differential equation:

\begin{equation}
{d \tau \over dr}=-\rho k {R_*^2\over r^2}.
\label{eqtau}
\end{equation}
\noindent

\noindent In order to close the system of equations we need to find the $f_{i}$'s, which,in turn, depend on the type and the amount of dust formed. In the present investigation, 
focused on the winds of O-rich AGB stars, we account for the formation of alumina dust, 
solid iron and Mg-silicates (forsterite, enstatite and quartz).
For alumina dust and Mg-silicates we consider the amorphous state since the physical conditions present in stellar outflows are more favourable to condensation as amorphous material than crystallized form. Clearly, we cannot rule out that part 
of the dust is present in crystalline state (e.g. corundum or crystalline Al$_{2}$O$_{3}$), which would imply the appearance of strong and specific solid-state features in the mid-infrared spectra of O-rich AGB stars; e.g. the spectral features of corundum at 13 $\mu m$ and of crystalline silicates at $\sim 10\  \mu m$ as well as at 20 $\mu m$.

We describe the dust growth process by vapour deposition on the surface
of  some pre-formed seed nuclei, assumed to be nanometer sized spheres. Each
dust species has a key element, which is the least abundant among the elements
necessary to form the corresponding dust aggregate.  The growth of the size of
dust grains ($a$) of a given species $i$ is determined via a competition between
a production term ($J^{gr}_i$),  associated to the deposition of new $i$ molecules
on already formed grains, and a  destruction factor ($J^{dec}_i$), proportional to
the vapor pressure of the key species $i$  on the solid state:

\begin{equation}
\label{graingrowth}
\frac{da_i}{dt}=V_{0,i}\Biggl(J^{gr}_i-J^{dec}_i\Biggr),
\end{equation}

\noindent where $V_{0,i}$ is the volume of the nominal molecule in the solid. Both
$J^{gr}_i$ and $J^{dec}_i$ are directly dependent on the value of the sticking
coefficient, $\alpha_{i}$, which varies according to the species considered.
From the dust grains size ($a_i$) we calculate the degree of condensation of the
key element into solid state ($f_{i}$) via the expression:

\begin{equation}
f_i=\frac{4\pi(a^3_i - a_{0,i}^3)}{3V_{0,i}}\frac{\epsilon_d}{\epsilon_k}
\label{f}
\end{equation}

\noindent where the initial dust grains size ($a_{0,i}$) is assumed to be equal to a
midrange value of 0.01 $\mu m$, $\epsilon_k$ is the number density of the key
elements in the wind, normalized to the  hydrogen density, and the normalized
density of the seed nuclei $\epsilon_d$ is assumed to be $10^{-13}$
\citep{knapp85}.

Finally, to determine the mass M$_i$ of the dust species $i$ produced during the
entire AGB phase, we use the equation:
 
\begin{equation}
\frac{dM_i}{dt}=\dot{M}X_{k}\frac{A_i}{A_k}f_i.
\label{masstot}
\end{equation}

\noindent where $X_k$ and $A_k$ are the surface mass fraction and molecular weight of the 
key-element, and $A_i$ is the molecular weight of the dust species considered.

\section{Alumina dust in the winds of AGBs}

The thermodynamical conditions of O-rich AGB winds are favorable to
form various types of oxygen-based dust, including alumina dust. This is because these stars evolve at
large luminosities, they lose mass at very high rates and their envelopes are extremely
cool; thus, the dust formation region is close to the stellar surface, where the
densities are sufficiently large to allow condensation into dust of 
considerable amounts of gas molecules.

In this section, we discuss the main properties of Al$_{2}$O$_{3}$ to understand its 
condensation process and the impact on the overall dust production in the winds of O-rich AGB stars. 

\subsection{Al$_{2}$O$_{3}$ properties}

The reaction leading to the formation of Al$_2$O$_3$ is

\begin{equation}
\label{reazAl2O3}
Al_2O + 2H_2O \rightleftharpoons Al_2O_3(s) + 2H_2
\end{equation}

where $Al_2 O$ and $H_2O$ are the most abundant Al- and O-based molecules not
bound in solid state\footnote{Indeed use of eq. 10 assumes that all the gaseous Al available 
is locked into Al$_{2}$O molecules. Actually AlOH is also expected to be present in not 
negligible quantities in AGB winds \citep{Sharp90}, which would require the treatment of 
the alternative channel for the formation of Al$_{2}$O$_{3}$, i.e. 
$2AlOH+H_2O \rightleftharpoons Al_2O_3(s) + 2H_2$. The results we obtain, neglecting the Al
locked into AlOH, might partly overestimate the amount of $Al_2O_3$ formed.}. 
Clearly, aluminium is the key-element for the production of this dust specie. 

\begin{figure}
\resizebox{1.\hsize}{!}{\includegraphics{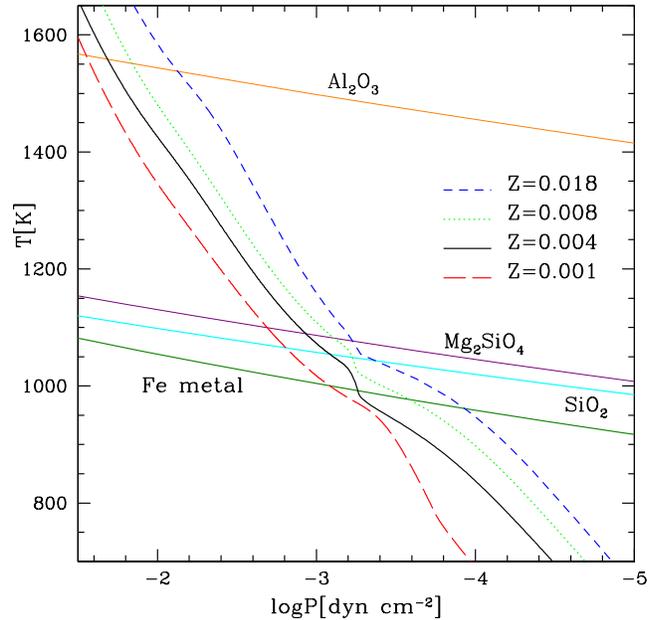}}
\vskip 30pt
\caption{Thermodynamic stratification in the P-T plane of the winds of 6 M$_{\odot}$ models in the phase of maximum luminosity when $\sim$1M$_{\odot}$ was lost from the envelope. The different lines correspond to the metallicities Z =
0.001 (red dashed-dotted line), Z = 0.004 (black solid line), Z = 0.008 (green dotted
line), and Z = 0.018 (blue dashed line).  The surface mass fraction of (O, Mg, Al, Si) for the models shown are: (6.95$\times 10^{-3}$, 7.2 $\times 10^{-4}$, 6.5 $\times 10^{-5}$, 7.9 $\times 10^{-4}$) for Z=0.018, (2.97$\times 10^{-3}$, 4.2 $\times 10^{-4}$, 2.7 $\times 10^{-5}$, 4.0 $\times 10^{-4}$) for Z=0.008, (1.29 $\times 10^{-3}$, 2.1 $\times 10^{-4}$, 1.9 $\times 10^{-5}$, 2.0 $\times 10^{-4}$) for Z=0.004 and (1.64 $\times 10^{-4}$, 4.8 $\times 10^{-5}$, 1.3 $\times 10^{-5}$, 5.6 $\times 10^{-5}$) for Z=0.001. We also show the stability curves in the pressure-temperature (P-T) plane for Al$_{2}$O$_{3}$ 
(orange), Mg$_2$SiO$_4$ (purple), SiO$_2$ (cyan) and iron (dark green).
}
\label{stability}
\end{figure}

To analyze the formation of 
alumina dust relatively to Mg-silicates, we show in Fig. \ref{stability} the stability lines of Al$_2$O$_3$, forsterite (Mg$_2$SiO$_4$), quartz (SiO$_2$) and iron, in the P-T plane. We clearly see that alumina, owing to an extremely high energy of formation \citep{Sharp90}, is by far the most stable compound, allowing the formation of Al$_{2}$O$_{3}$ grains at temperature as high as $\sim$ 1500 K. In the same Figure,
we also show the structure of the winds surrounding models of a 6 $M_{\odot}$ initial mass at different metallicities. The plot refers to the phase of highest luminosity, with the maximum strength of the dust formation process, after $\sim$ 1 M$_{\odot}$ was lost by the star.

The refraction index used for the computation of the extinction coefficients for alumina dust are
taken from \citet{Koike95}, who present optical constants for crystalline Al$_{2}$O$_{3}$.
A comparison with the more recent results from 
\citet{Begemann97}, who consider the amorphous case, indicates no meaningful differences 
in the Al$_{2}$O$_{3}$ production. This adds more robustness to our
results. 

However, concerning the amorphous or crystalline nature of the Al$_{2}$O$_{3}$ formed,
any prediction is made difficult by the fact that the threshold temperature above which
the crystalline component is dominant \citep[$\sim 1440$K][]{levin98} is within the range
of temperatures at which the condensation process takes place (1200--1500K). The high abundance
of amorphous alumina in the circumstellar envelopes of many O--rich AGB stars confirms the
presence of the amorphous component, but the details of the relative distribution of the
two phases of Al$_{2}$O$_{3}$ is beyond the scopes and the possibilities of the present
analysis.
From our analysis, however, we cannot draw any conclusion on the amorphous 
or crystalline nature of the Al$_{2}$O$_{3}$ formed.

The rate at which Al$_{2}$O$_{3}$ grains grow in the expanding winds of AGBs is
unfortunately made uncertain by the poorly known sticking coefficient. To date, the only robust measurement, limited to the crystalline form, indicates a value for the $\alpha_{al}$ smaller than 0.1
\citep{takigawa12}. In analogy with the amorphous Mg-silicates, we thus assume as reference value $\alpha_{al}$=0.1.
However, we investigate the sensitivity of our results to
$\alpha_{al}$ in Section 4.2, where we explore different values
of the sticking coefficient.

\subsection{Hot bottom burning in massive AGBs and the activation of the Mg-Al
chain} 

As we have mentioned above, stars of initial mass above $\sim 3$M$_{\odot}$
(this limit depends also on the convection model adopted, see 
Renzini \& Voli 1981, Ventura \& D'Antona 2005, Garc\'{\i}a-Hern\'andez et al. 2013) 
experience hot bottom burning. The bottom of the convective envelope becomes sufficiently
hot to activate an advanced proton--capture nucleosynthesis, with the consequent
modification of the surface chemistry. The temperature at the bottom of the
convective envelope T$_{\rm bce}$ is the key quantity in determining the
extent of the nucleosynthesis experienced.

\begin{figure*}
\begin{minipage}{0.45\textwidth}
\resizebox{1.\hsize}{!}{\includegraphics{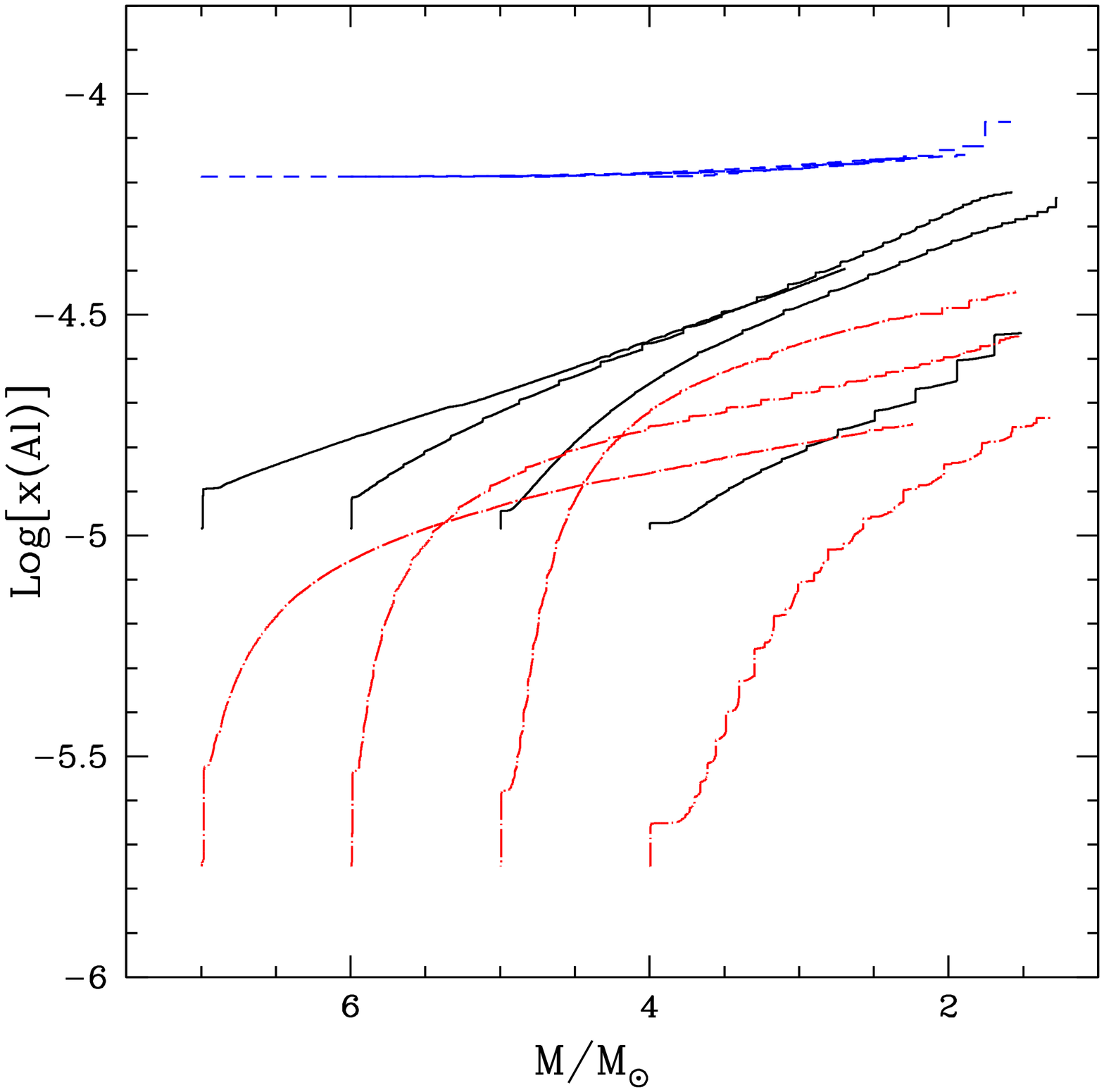}}
\end{minipage}
\begin{minipage}{0.45\textwidth}
\resizebox{1.\hsize}{!}{\includegraphics{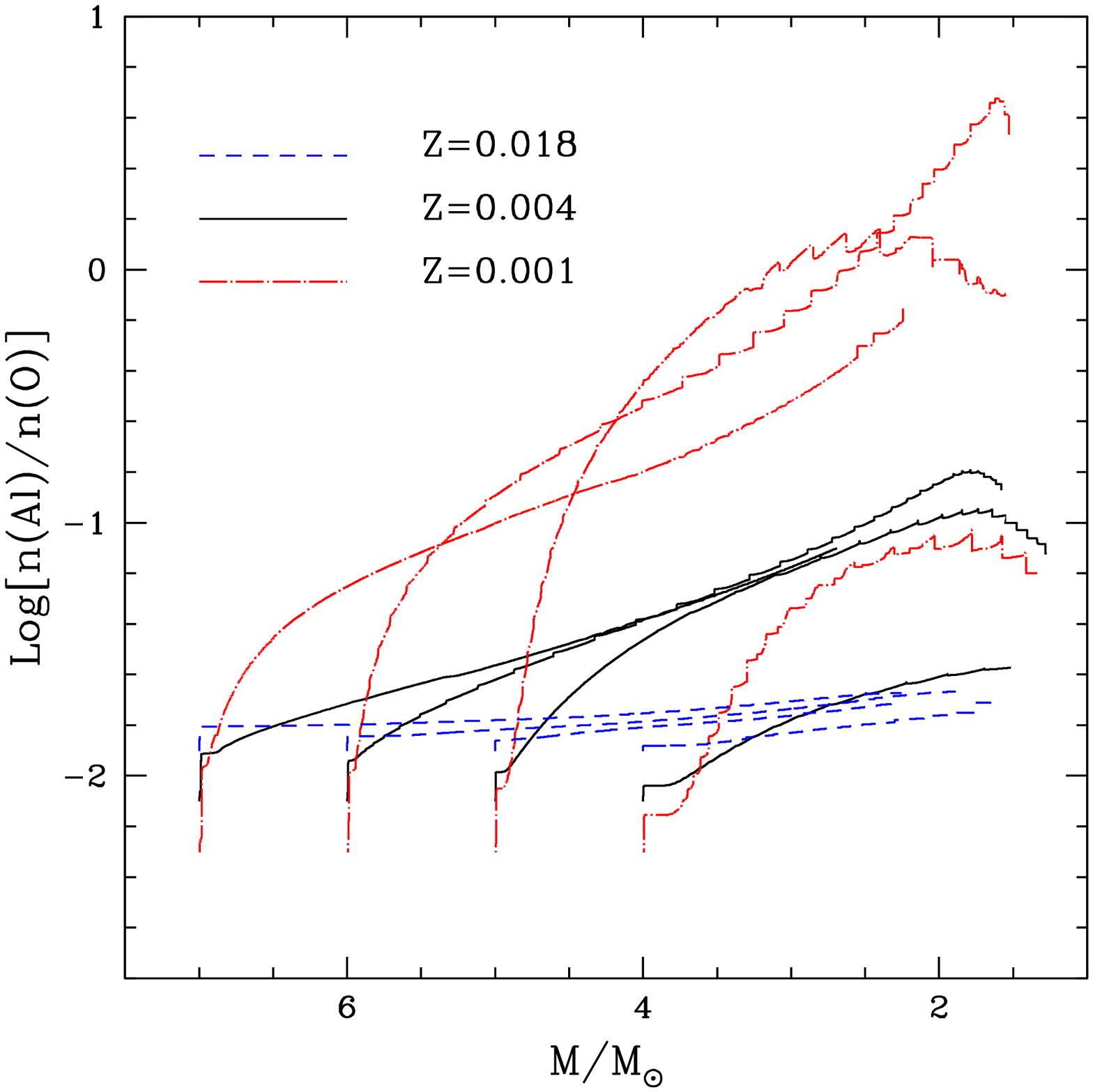}}
\end{minipage}
\vskip 30pt
\caption{ The surface mass fraction abundance of Al (left panel) and the number density ratio
of Al and O (right panel) as a function of the current mass of the star.
The predictions for models of 4, 5, 6, and 7 $M_{\odot}$ at Z = 0.001 (red
dashed-dotted line), 0.004 (black solid line) and
0.018 (blue dashed line) are displayed. Note that stellar mass decreases during
the evolution in the AGB and takes place from left to right.}
\label{xal}
\end{figure*}

T$_{\rm bce}$ generally increases with core mass, and is higher for lower stellar
metallicities model \citep{ventura13}. When T$_{\rm bce}$ reaches $\sim
30$ MK, lithium production via the Cameron--Fowler mechanism and carbon
destruction via p-capture are achieved. At temperatures of the order of $\sim
70-80$ MK, oxygen undergoes proton fusion and is destroyed.
When T$_{\rm bce}$ exceeds $\sim 80$ MK, the innermost regions of the envelope
become site of a series of proton capture reactions involving the isotopes of
magnesium, that eventually lead to the production of aluminium \citep{vcd11}.
Because the initial magnesium is much larger than aluminium, even a small
depletion of the surface magnesium is sufficient to induce a considerable
increase in the surface Al--content. 

Ignition of the Mg-Al nucleosynthesis is crucial for the discussion concerning
the production of Al$_{2}$O$_{3}$, because in the vast majority of cases
aluminium drives the formation process for this species.

The left panel of Figure \ref{xal} shows the variation of the surface
Al-content of models of different mass, for three of the four metallicities investigated
in this work (for clarity reasons, in the figure we omit Z=0.008). The choice of the logarithmic scale allows to appreciate the
extent of the Al-increase.

In agreement with the previous discussion, we note the following:

\begin{enumerate}

\item{The percentage increase in the surface aluminium is higher in models
of lower metallicity. While in the Z$=10^{-3}$ case the surface Al is increased
by a factor $\sim 10-20$, in the Z$=8\times 10^{-3}$ the increase is limited to a 
factor $\sim 2$. No change is found in the solar case.}

\item{Models of higher mass generally produce more aluminium, because they
experience a  stronger HBB. However, the most massive models of metallicity Z$=10^{-3}$
and Z$=4\times 10^{-3}$ eject gas less enriched in aluminium than their smaller
mass counterparts; this is due to the strong mass loss experienced, so that
the envelope is completely lost before a great production of aluminium
occurs \citep{vcd11}.}

\end{enumerate}

Figure \ref{xal} (right panel) shows the variation of Al/O in the same
models shown in the left panel. Al/O increases with time due to the
simultaneous production of aluminium and destruction of oxygen, but remains below
unity in all cases, with the only exception of the latest evolutionary phases of the more 
massive models at Z$=10^{-3}$: only in these latter models the destruction of the surface 
oxygen eventually leads to the condition Al/O$>1$, which makes oxygen 
the key element in the production of Al$_{2}$O$_{3}$. Therefore, we
may safely assume in the following analysis that aluminium is the key--element for the 
formation of alumina dust.

\section{Alumina dust production}
Here, we discuss dust formation in AGB models of metallicities $3\times 10^{-4} \leq Z \leq 0.018$.
Because Al$_{2}$O$_{3}$ forms only in O--rich environments, we restrict the present
analysis to stars of mass above 3$M_{\odot}$. Models with $Z=4\times 10^{-3}$
and $Z=0.018$ have been calculated specificaly for this paper.
Models with $Z=3\times 10^{-4}$, $10^{-3}$, $8\times 10^{-3}$ were published 
in previous investigations by our group \citep{paperI, paperII, Dic13, paperIV}. However,
the dust formation modelling was repeated here, because Al$_{2}$O$_{3}$ formation was ignored in
our previous works.  
In addition, we may disregard the Z$=3\times 10^{-4}$ models 
from our investigation - dust production in O-rich AGB stars of this metallicity is too low to
drive the wind, owing to the extremely low abundances of silicon and aluminium (see  \citet{Dic13}). 

\begin{figure}
\resizebox{1.\hsize}{!}{\includegraphics{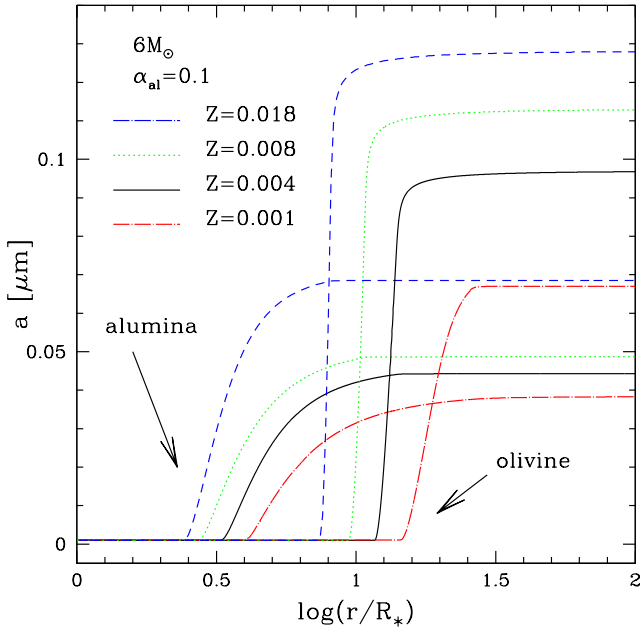}}
\vskip 30pt
\caption{Size of the Al$_{2}$O$_{3}$ and forsterite dust grains as a function
of the distance from the centre of the star, for the same models of Fig. \ref{stability}. Note that we have assumed the sticking
coefficient $\alpha_{al}=0.1$ in the simulations.}
\label{rad}
\end{figure}

\subsection{The growth of Al$_{2}$O$_{3}$ grains}
Figure \ref{rad} shows the size of Al$_{2}$O$_{3}$ and Mg-silicates (here represented by forsterite dust)
grains at different distances from the centre of the star, for the same models of Fig. \ref{stability}. Iron grains form in minor quantities, in even more external regions, compared to  Mg-silicates. They are omitted, for clarity reasons, in the present and the following figures.
These profiles refer to an interpulse phase during the AGB evolution, when the
dust production is at the highest rate; this occurs in all the cases shown 
in Figure \ref{rad} after $\sim 1$M$_{\odot}$ was lost from the star.

The exact location of the condensation zone is mainly determined by the effective temperature 
($T_{eff}$) of the star (see eq. \ref{eqteff}). Therefore, the trend with metallicity is straightforward: in solar metallicity models, owing to their 
lower $T_{eff}$, the growth of Al$_{2}$O$_{3}$ particles begins at $\sim 2 R_*$ from the center of the star (blue dashed track in Fig.\ref{rad}), 
while at $Z=0.001$ the Al$_{2}$O$_{3}$ condensation zone is in more external circumstellar regions, at $\sim 4 R_*$ (red dotted-dashed line). On the other hand, a change in 
the initial mass of the star does not strongly affects these results because models with the same $Z$ and different masses evolve at approximately
the same $T_{eff}$.
 
The condensation of Al$_{2}$O$_{3}$ does not inhibit (or severely influence) the formation of Mg-silicates, owing to its large transparency: the acceleration of the wind via radiation pressure starts further the formation of the alumina dust, where Mg-silicates begins to grow. Although, the formation of alumina dust determines a slight increase in the opacity, in turn, leads to a steeper gradient of the optical depth
($\tau$, see eq. \ref{eqtau}). Because of the boundary condition that $\tau$ vanishes at infinity, this can be accomplished only via an increase of $\tau$ in the regions internal to the Al$_2$O$_3$ formation layer. The higher $\tau$ favoures an increase of the temperatures (see eq. \ref{eqteff}). Therefore, in the present models, forsterite dust grains begin to grow in more
external regions, at a distance of $\sim 10 {\rm R}_{\ast}$ from the star's centre, where the densities are smaller and the amount of forsterite dust produced
is thus consequently slightely reduced. This is accompanied by a larger production of enstatite and quartz dust. The total effect on the amount of dust product it will be discuss in Section 4.3.

\begin{figure*}
\begin{minipage}{0.33\textwidth}
\resizebox{1.\hsize}{!}{\includegraphics{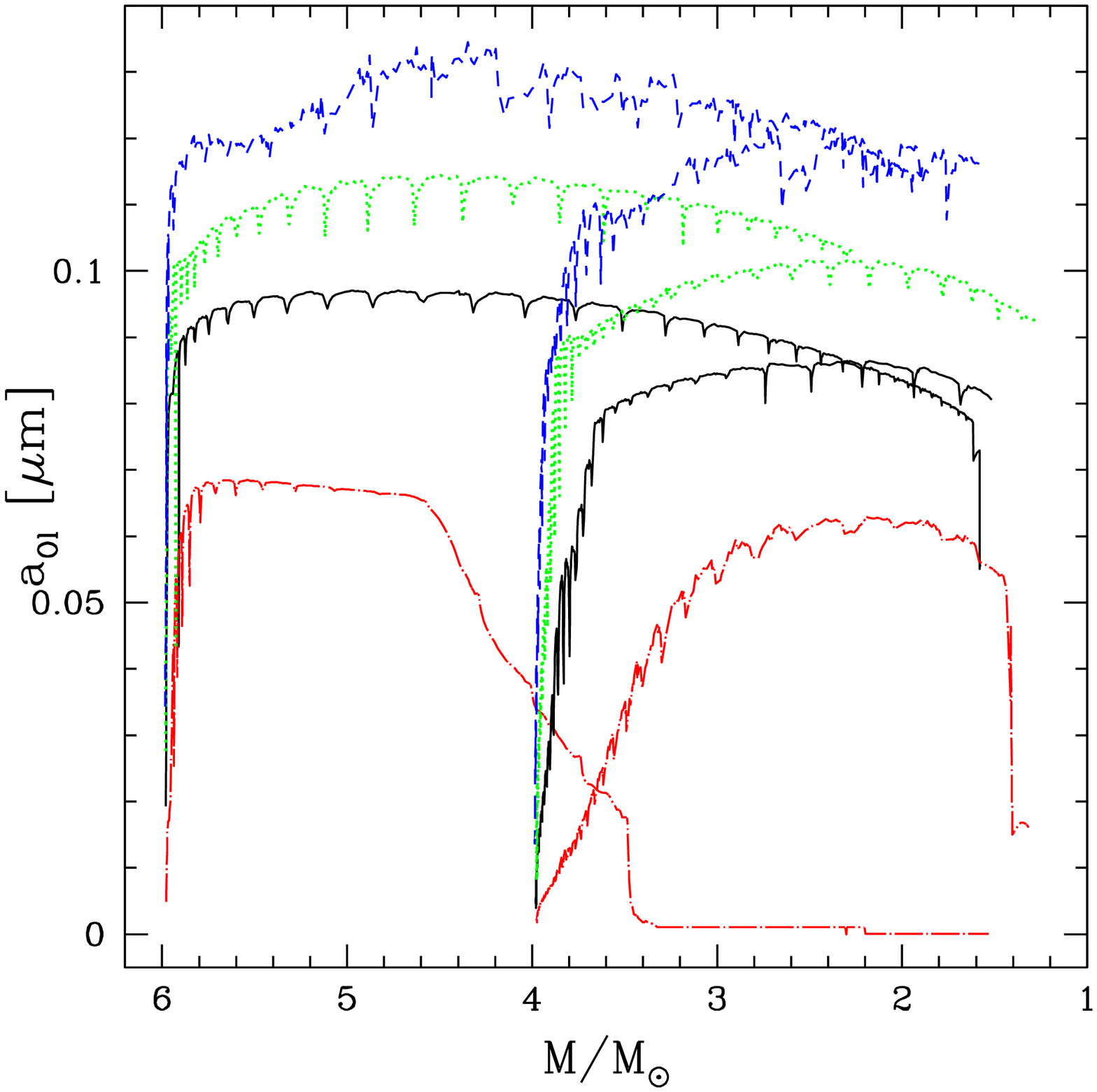}}
\end{minipage}
\begin{minipage}{0.33\textwidth}
\resizebox{1.\hsize}{!}{\includegraphics{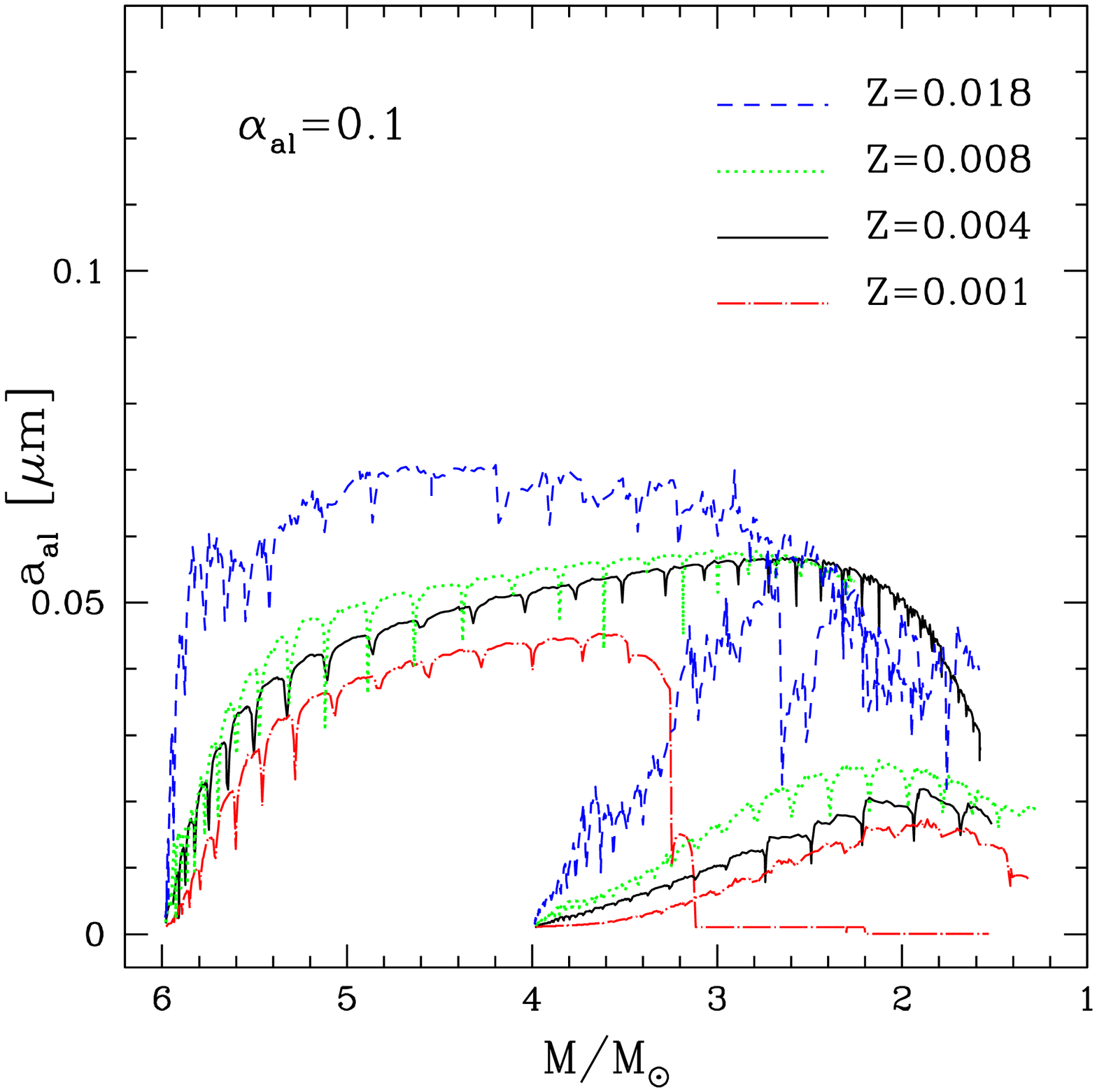}}
\end{minipage}
\begin{minipage}{0.33\textwidth}
\resizebox{1.\hsize}{!}{\includegraphics{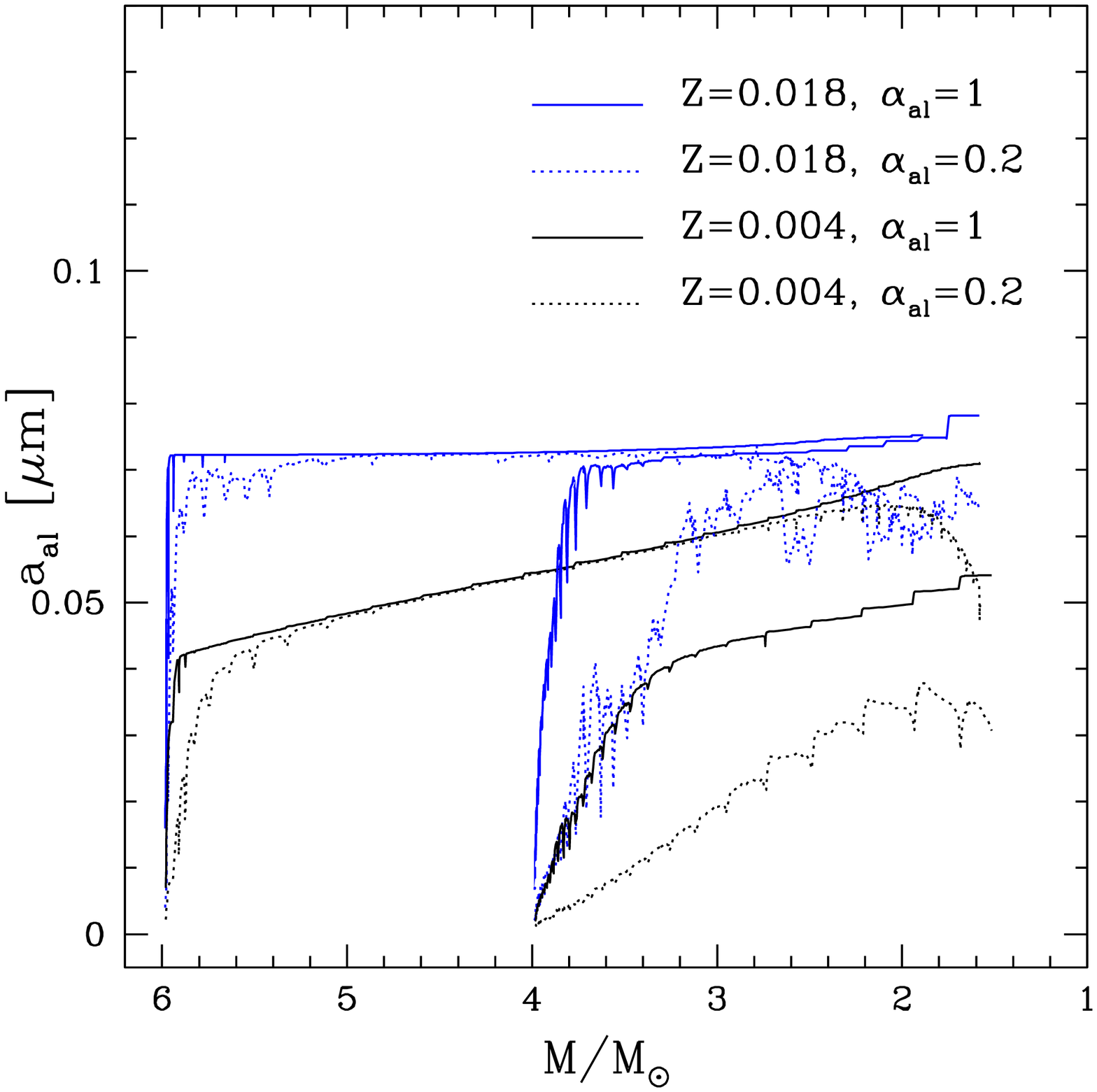}}
\end{minipage}
\vskip 30pt
\caption{Evolution as a function of the total mass of the star
 of the forsterite (left panel) and
Al$_{2}$O$_{3}$ (middle panel) dust grain sizes for models with progenitor
masses of 4 and 6 $M_{\odot}$ at Z = 0.001 (red dashed-dotted line), 0.004
(black solid line), 0.008 (green dotted line), and 0.018 (blue dashed line), assuming $\alpha_{al}=0.1$. In
the right panel we display the Al$_{2}$O$_{3}$ dust grain size versus the
evolution of the stellar mass for models with initial masses of 4 and 6
$M_{\odot}$ models, at  Z = 0.004 (black) and Z = 0.018 (blue), and by assuming two different values of the sticking
coefficient - $\alpha_{al}$ = 0.2 and 1 (dotted and solid lines, respectively).}
\label{dimension}
\end{figure*}

\begin{figure*}
\begin{minipage}{0.45\textwidth}
\resizebox{1.\hsize}{!}{\includegraphics{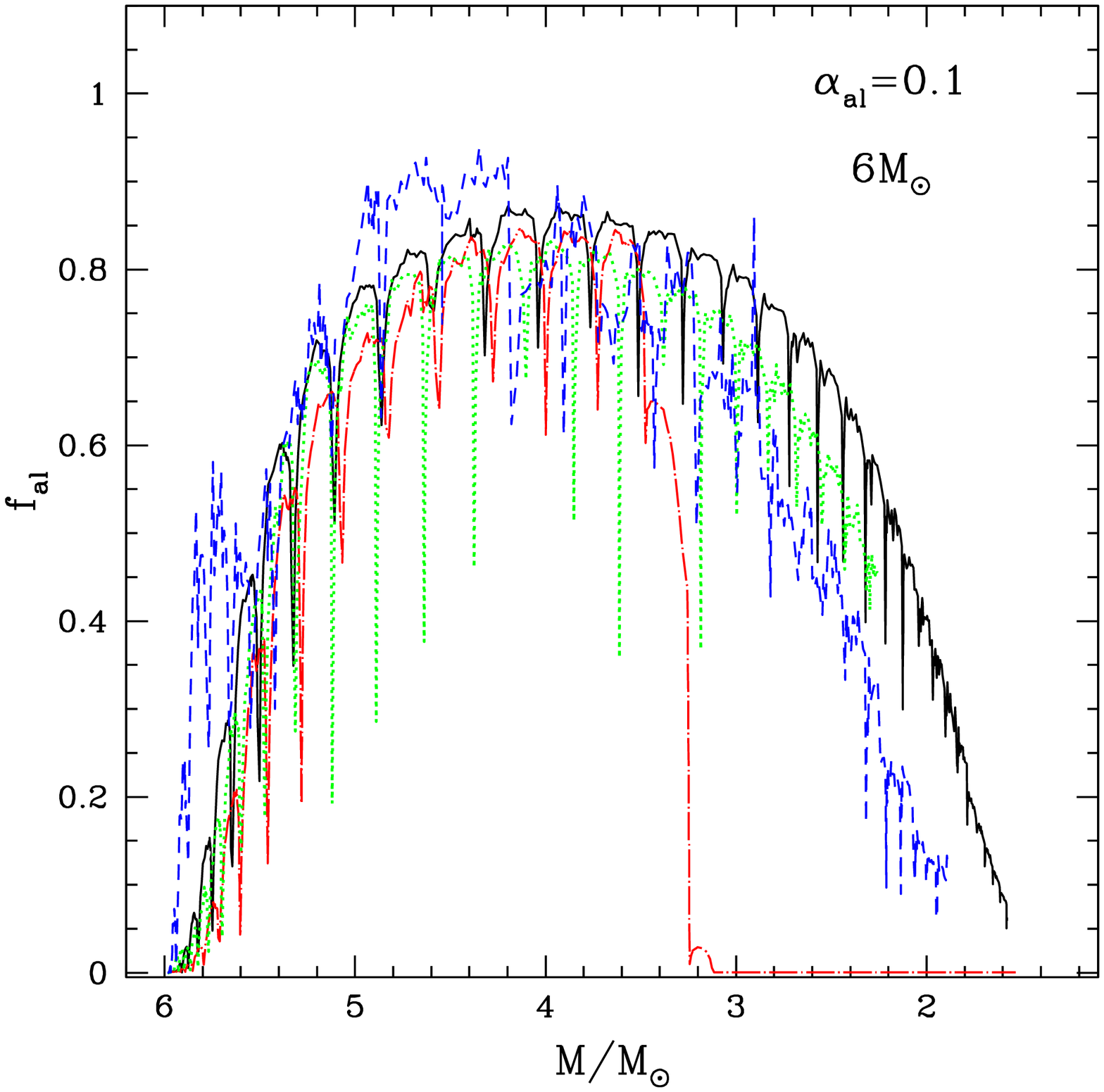}}
\end{minipage}
\begin{minipage}{0.45\textwidth}
\resizebox{1.\hsize}{!}{\includegraphics{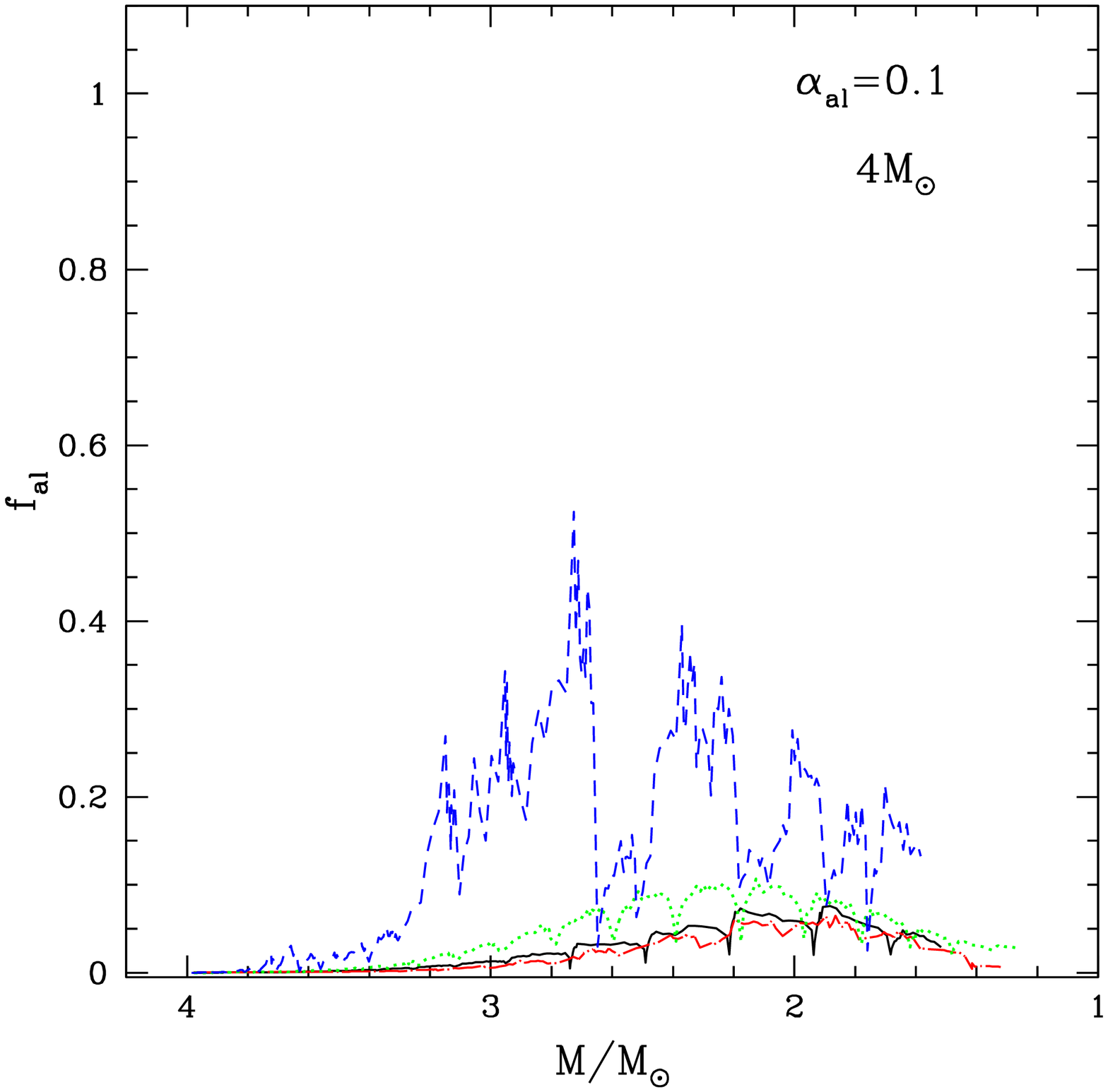}}
\end{minipage}
\vskip 30pt
\caption{Evolution of the aluminium fraction condensed in Al$_{2}$O$_{3}$
($f_{al}$) as a function of the evolution of the
total mass by assuming $\alpha_{al}=0.1$.  We consider 6 $M_{\odot}$ (left panel) 
and 4 $M_{\odot}$ (right panel) initial mass model, at Z =
0.001 (red dashed-dotted line), Z = 0.004 (black solid line), Z = 0.008 (green dotted
line), and Z = 0.018 (blue dashed line).}
\label{fco}
\end{figure*}

The left and middle panels of Figure \ref{dimension} show the variation during the AGB phase 
of the size of forsterite and Al$_{2}$O$_{3}$ grains, respectively. The different lines correspond
to models of initial masses 4 and 6 $M_{\odot}$ of several metallicities 
(other masses were omitted for clarity reasons). 

The two panels of Fig. \ref{fco} show, for the same models, the variation of $f_{al}$ (see Eq. \ref{f}), the fraction of 
aluminium condensed into dust.

The results obtained can be synthesized as follows:

\begin{enumerate}

\item{The size reached by forsterite and Al$_{2}$O$_{3}$ grains increases with
metallicity, owing to the larger silicon and aluminium mass
fractions in the surface layers of higher--Z models.}

\item{For a given Z, the dust grain size increases with the
progenitor mass. This is because more massive stars experience a
stronger HBB, and thus  they evolve at higher luminosities and experience higher
mass-loss rates.}

\item{The grain size evolution of forsterite and Al$_{2}$O$_{3}$ dust during
the AGB phase is rather different. The maximum size of forsterite particles is reached when less than half of the envelope mass is lost. The dimension of the forsterite grains depends on metallicity, ranging from $\sim 0.07 \mu m$ for $Z= 10^{-3}$, to $0.13 \mu m$ for solar chemistry.
Conversely, the size of Al$_{2}$O$_{3}$ grains increases during almost the whole AGB evolution. The decrease in the forsterite grain
size is due to the drop in the total luminosity (and hence of the mass-loss
rate) when the envelope mass is consumed. In the case of alumina dust,
this effect is counterbalanced by the gradual
increase in the surface Al content, favored by the
activation of the Mg-Al nucleosynthesis. This is particularly important in 
massive, metal-poor AGB models (see left panel of Figure \ref{xal}).}

\item{Solar metallicity models behave somewhat differently from their
lower Z counterparts: the Al$_{2}$O$_{3}$ grain size attains the maximum
values of $a_{al} \sim 0.075 \mu m$ since the early AGB phases. 
For models of subsolar chemistry, $a_{al}$ gradually increases as the star evolves, until a maximum value of $0.05 - 0.06 \mu m$ (slightly dependent on $Z$) is reached in the more massive AGBs (see central panel of Fig.\ref{dimension}). 
The alumina dust grains of the largest size are thus expected to form in the
circumstellar shells of massive AGBs of solar metallicity, with
$a_{al} \sim 0.07-0.08 \mu$m. }

\item{A high percentage of gaseous aluminium condenses into Al$_{2}$O$_{3}$
(almost $\sim 90 \%$ in the higher mass AGB models, see left panel of Figure
\ref{fco}). We note that the evolution of $f_{al}$ does not closely follow
that of $a_{al}$ (the Al$_{2}$O$_{3}$ dust grain size) owing to the increase in
the surface aluminium abundance (see Eq. \ref{f}).}

\item{Although Al$_{2}$O$_{3}$ dust grains form in more internal circumstellar
regions and a high fraction of aluminium is condensed into dust, the size of the forsterite grains is still larger than that of Al$_{2}$O$_{3}$. This is because
the amount of silicon available is always much larger than aluminium.}

\end{enumerate}

The results given in points (i) (iii) and (iv) above are partly dependent on the 
choice of the initial density of seed grains, $\epsilon_d$. To understand how
critical the choice of $\epsilon_d$ is, we run some simulations where $\epsilon_d$
was increased/decreased by a factor 10. The results in terms of the size reached
by the grains of the various species of dust showed up only a modest dependence
on $\epsilon_d$, with a maximum variation by a factor 2 for a 1 dex variation of
$\epsilon_d$. The reason for this is that, based on Eq.8, the fraction $f$ of the
key--species condensed into dust goes as $\sim a^3 \epsilon_d$. Because the number
of gaseous molecules available (that determines the growth rate of the dust grains)
depends critically on $f$, ($n \sim (1-f)$), the effect of
increasing/decreasing $\epsilon_d$ is partly counterbalanced by the same
decrease/increase in $a^3$. Note that a linear relation between $\epsilon_d$ and Z was invoked to account for the larger availability of the seed nuclei in more metal-rich environments \citep{nanni}. Based on the arguments given above, such a scaling relation would reduce the difference in the size of Mg-silicates and alumina dust grains formed around models of different metallicity and would leave the initial mass of the star as the dominant factor determining the dimension of the particles formed.

\begin{figure*}
\begin{minipage}{0.45\textwidth}
\resizebox{1.\hsize}{!}{\includegraphics{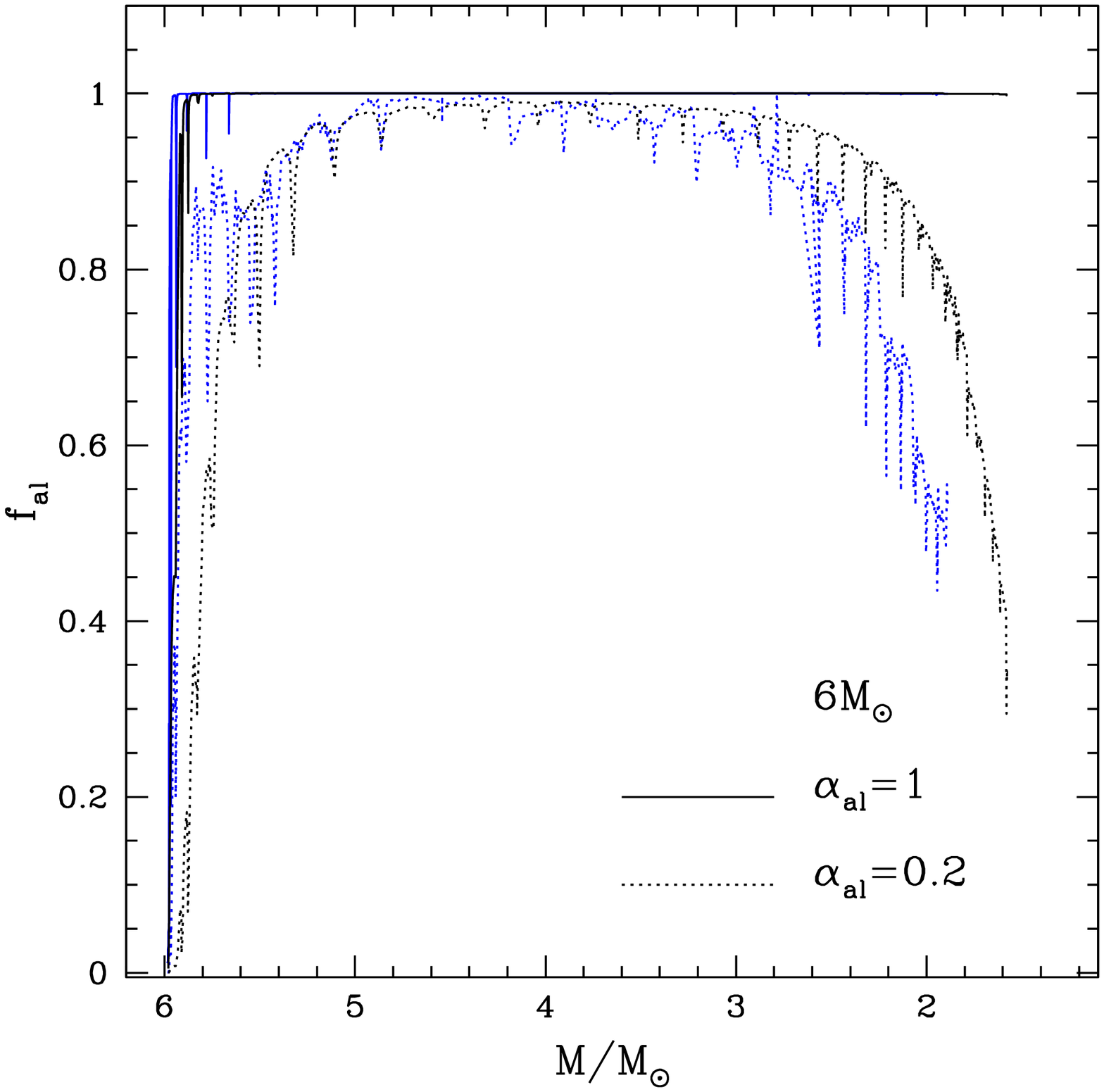}}
\end{minipage}
\begin{minipage}{0.45\textwidth}
\resizebox{1.\hsize}{!}{\includegraphics{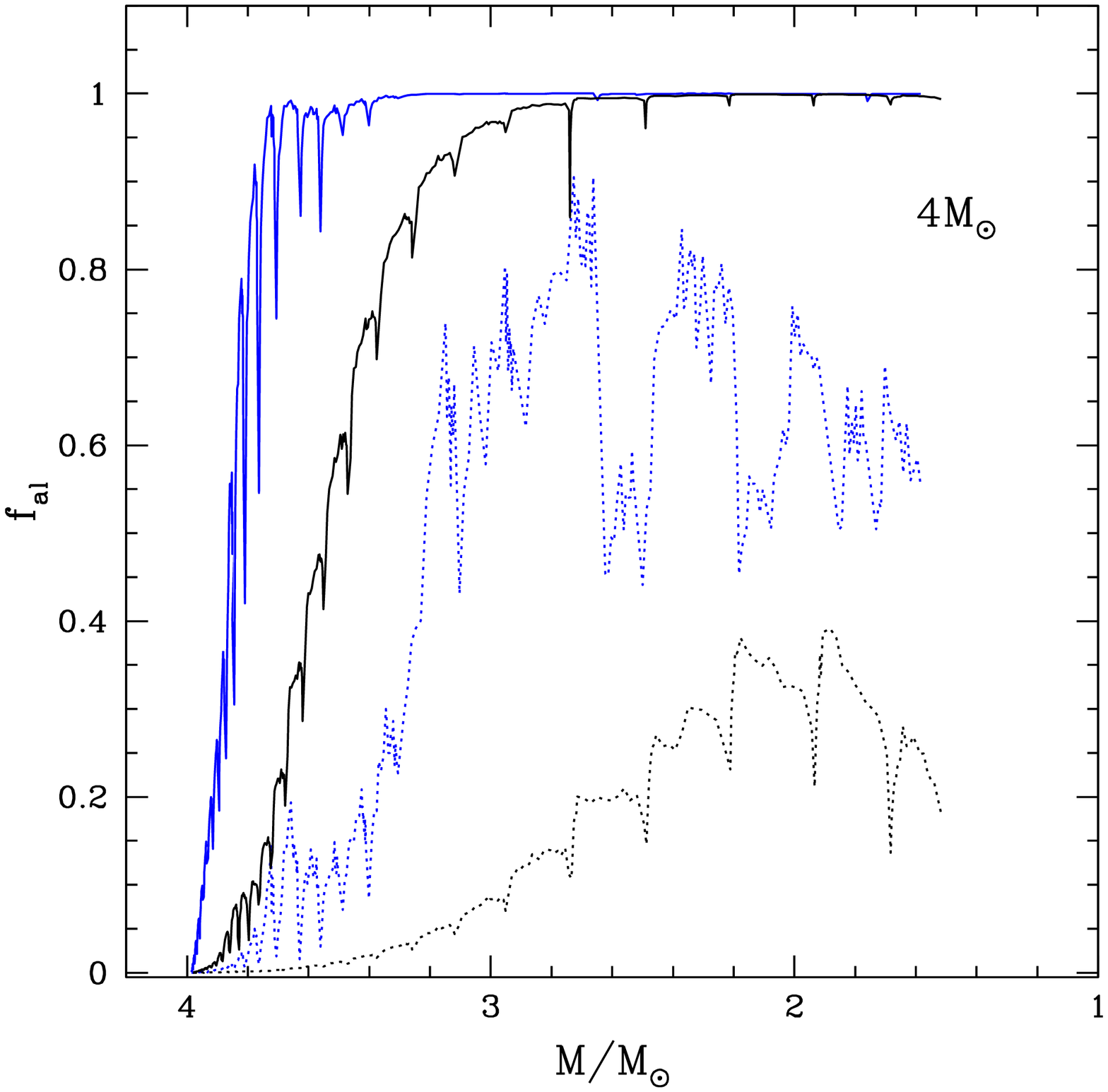}}
\end{minipage}
\vskip 40pt
\caption{Evolution of the aluminium fraction condensed in Al$_{2}$O$_{3}$
($f_{al}$) as a function of the evolution of the
total mass for 6 $M_{\odot}$ (left panel) 
and 4 $M_{\odot}$ (right panel) initial mass model, at Z = 0.004 (black) and Z = 0.018 (blue).  We explore $\alpha_{al}$ =
0.2 (dotted line) and $\alpha_{al}$ =1 (solid lines) for both cases.}
\label{falfco}
\end{figure*}

\subsection{The role of the sticking coefficient}
Given the poor knowledge of the Al$_{2}$O$_{3}$ sticking
coefficient ($\alpha_{al}$) we discuss how the results presented in the previous
sections depend on the choice of $\alpha_{al}$, by exploring different 
values until $ \alpha_{al}$=1.

We focus on the 4M$_{\odot}$ and 6M$_{\odot}$ models of metallicity
$Z=4\times 10^{-3}$ and $Z=0.018$. The 
evolution of the Al$_{2}$O$_{3}$ dust grain sizes obtained when using
$\alpha_{al}=0.2$ and $\alpha_{al}=1$ are shown in the right panel of Figure
\ref{dimension}, while the corresponding evolution of the fraction of  aluminium
condensed into Al$_{2}$O$_{3}$ is displayed in Figure
\ref{falfco}. 

As expected, adopting a larger $\alpha_{al}$ increases the
Al$_{2}$O$_{3}$ grain size. However, in the 6$M_{\odot}$ models of both metallicities $a_{al}$ 
is almost independent of $\alpha_{al}$: this is because $f_{al}$ attains very large values, close to 
unity, even for $\alpha_{al}=0.1$ (see left panel of Figure \ref{fco}). Under these conditions, a further 
increase in $\alpha_{al}$ hardly leads to a further growth of the Al$_{2}$O$_{3}$ particles, which 
would be possible only if a considerable increase in the surface Al occurred. 

Lower mass models behave somewhat differently. Unlike their more
massive counterparts, for both the cases $Z=4\times 10^{-3}$ and $Z=0.018$
we note a higher sensitivity to $\alpha_{al}$. This stems from the fact that
the saturation conditions are never reached, with $f_{al}$ evolving below
$\sim 0.5$ for $\alpha_{al}=0.1$ (see right panel of Figure \ref{fco}).
Indeed, the right panel of Figure \ref{falfco} shows that saturation in the
4M$_{\odot}$ case is only reached for $\alpha_{al}=1$. Note that for this mass,
the maximum size reached by the Al$_{2}$O$_{3}$ grains - in the $\alpha_{al}=1$ case 
for the two metallicities mentioned above - is a$_{al}$ $\sim$ 0.075 $\mu$m, which
adds more robustness to the conclusions given in the previous subsection.

In case of an extremely small value of $\alpha_{al}(=0.01)$ the growth of Al$_{2}$O$_{3}$ grains is severely inhibited. In the most massive models, suffering the strongest HBB, this would reflect in a decrease in the size of Al$_{2}$O$_{3}$ particles by a factor $\sim 2-3$. In the M-stars of smaller mass, such a small $\alpha_{al}$ would strongly suppress the formation of Al$_{2}$O$_{3}$, whose grains would hardly exceed nanometer size dimensions.

\subsection{Al$_{2}$O$_{3}$ mass production}
\begin{figure*}
\resizebox{1.\hsize}{!}{\includegraphics{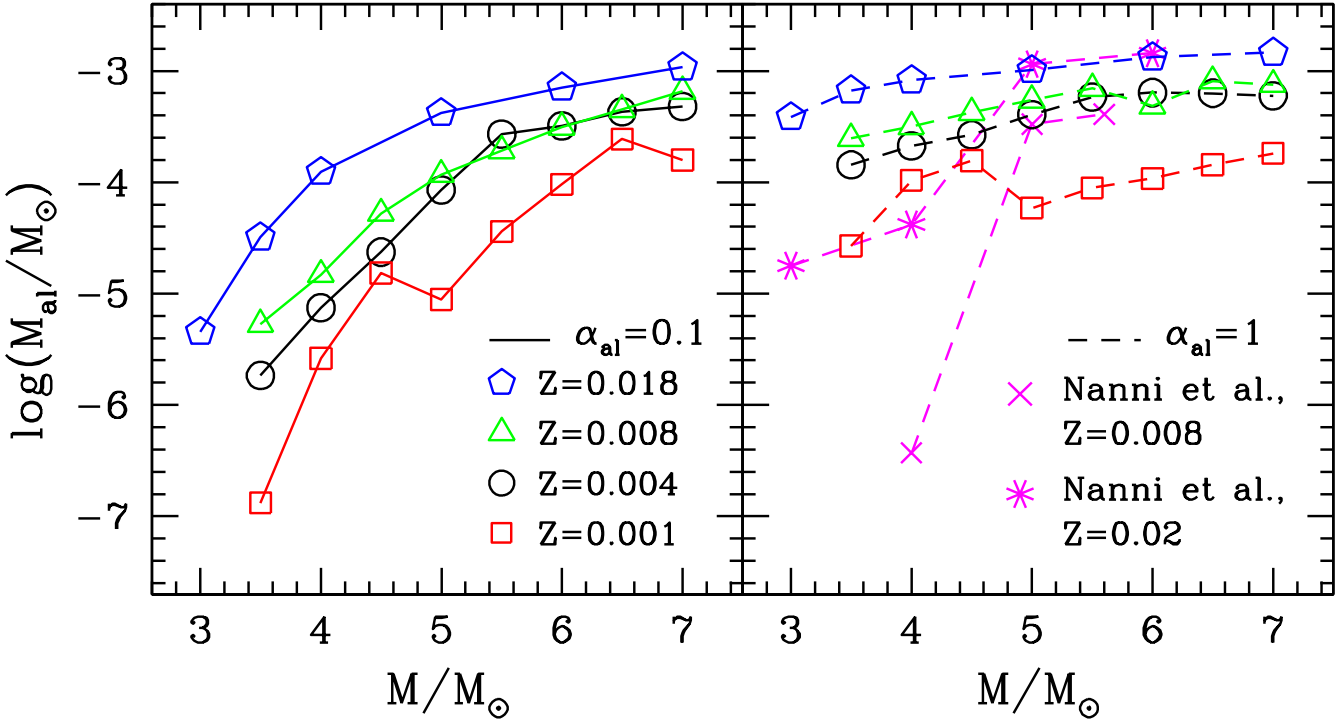}}
\vskip 40pt
\caption{Alumina dust mass produced during the entire AGB phase as a function of
the initial stellar mass at Z = 0.001 (red open squares), 0.004 (black open circles),
0.008 (green open triangles), and 0.018 (blue open pentagons). In the left panel, we 
show results obtained by assuming $\alpha_{al}$ =0.1 (solid line), while in the right panel 
$\alpha_{al}$ is 1 (dashed line). In the right panel, we also report results from Nanni et 
al. 2013, which consider $\alpha_{al}$=1 at  Z=0.008 (magenta cross) and Z=0.02 (magenta star)}
\label{mass}
\end{figure*}

The overall mass production of alumina dust is calculated by means of 
equation \ref{masstot}. In the left panel of Figure \ref{mass}, we show the total mass of
Al$_{2}$O$_{3}$ produced ($M_{al}$) during the whole AGB phase for different
metallicities as a function of the initial mass of the star ($M$), assuming $\alpha_{al}=0.1$. 

The largest amount of alumina dust is produced by solar metallicity AGBs. $M_{al}$ is strongly dependent on $M$, ranging from $10^ {-5}M_{\odot}$ for $M=3M_{\odot}$,
to $10^ {-3}M_{\odot}$ for $M=7M_{\odot}$. This trend with the mass of the star is
found also for the other metallicities; $M_{al}$ scales approximately linearly with $Z$.

To have an idea of the uncertainties associated with the choice of the sticking coefficient, we
compare the results obtained with $\alpha_{al}=0.1$ with those for $\alpha_{al}=1$ (right panel of Fig. \ref{mass}). In the latter case, the trend of $M_{al}$ with the stellar mass is much
flatter. This is because the saturation conditions are reached even for the
lowest mass models experiencing HBB. In this case, the mass of alumina dust produced becomes 
practically independent of $M$, and scales approximately linearly with metallicity.

The comparison with the results by \citet{paperI, paperII}, where formation of
Al$_{2}$O$_{3}$ was not considered, allows to quantify the effects on the amount of Mg-silicates
formed.

The total mass of dust produced, $M_d$, increases when
the formation of Al$_{2}$O$_{3}$ is taken into account. For the more massive AGBs, $M_d$ 
increases by $\sim 6-7\%$, while the amount of Mg-silicates formed, $M_{\rm sil}$, decreases by 
$\sim 5\%$. For stars of lower mass no meaningful differences are found among the two cases,
owing to the small quantities of alumina dust formed.

The differences introduced by considering the formation of Al$_{2}$O$_{3}$ are larger
in the case $\alpha_{al}=1$, because the condensation process is more efficient.
In the massive AGBs domain the difference is purely quantitative, the total mass formed
being increased by $\sim 15\%$. Unlike the standard case, for $M\sim 3-4M_{\odot}$, $\sim 20\%$
of dust is under the form of Al$_{2}$O$_{3}$: neglecting the formation of alumina dust would
underestimate considerably the overall amount of dust formed.

 The reliability of the results obtained in terms of mass of dust $M_d$ produced is 
partly affected by the intrinsic indetermination in the choice of the density of seed
particles, $\epsilon_d$. However, as found for the dimension reached by the grains of the
various species, the sensitivity of $M_d$ on $\epsilon_d$ is modest, with a total
variation below $\sim 50\%$ for a variation of $\epsilon_d$ of one order of magnitude
(see eq.9 and the discussion at the end of section 4.1.

Very recently, \citet{nanni} presented models of dust formation 
around AGB stars, including also Al$_{2}$O$_{3}$ production, at three 
metallicities (Z = 0.001, 0.008, and 0.02) and assuming $\alpha_{al}=1$. 
Their models show no Al$_{2}$O$_{3}$ production at the lowest metallicity of 
0.001 and for progenitor masses below 5 M$_{\odot}$ at higher metallicity 
(see right panel of Figure \ref{mass}). This is in contrast with our models where we 
find, at least for the more massive ( $>$ 5 M$_{\odot}$) 
stars, that a significant amount of  Al$_{2}$O$_{3}$ is also formed at very low 
metallicity (Z = 0.001). This difference is mainly due to the different 
treatment of convection in the stellar evolution model. As we have 
mentioned above, the FST description of convection used in our models 
implies strong HBB conditions, which are not found in the \citet{nanni}
 AGB models. At the higher metallicities, Z = 0.008 and Z = 
0.02, and for initial masses $\ge$ 5 M$_{\odot}$, the  Al$_{2}$O$_{3}$ production found by 
\citet{nanni} is comparable to ours. Much larger differences are seen for the lower masses ($<$ 5 
M$_{\odot}$), in which the mass of  Al$_{2}$O$_{3}$ dust formed is always 
below 10$^{-4}$ M$_{\odot}$ in the \citet{nanni} models; on the contrary, 
our AGB models predict  Al$_{2}$O$_{3}$ production between 10$^{-4}$ and 10$^{-3}$ 
M$_{\odot}$. This difference is again due to the much softer HBB 
experienced by the Nanni et al. (2013) AGB models, where the lower mass 
($<$ 5 M$_{\odot}$) stars become C-rich, inhibiting the  Al$_{2}$O$_{3}$ production.

\section{Comparison with observations}

\subsection{Al depletion in HBB AGB stars: a further indicator of  Al$_{2}$O$_{3}$ dust production?}

From the previous sections we know that formation of Al$_{2}$O$_{3}$ grains is 
favored by HBB conditions. In addition, the production of alumina dust 
implies an important decrease in the abundance of gaseous aluminium in 
the AGB wind. An important consequence of the high aluminium fraction 
condensed in Al$_{2}$O$_{3}$ that we find in our models (especially in the more 
massive AGB models) is that we predict gaseous Al to be underabundant 
in the more massive HBB AGB stars. Interestingly, \citet{mcsaveney} found gaseous Al to be severely depleted (by almost one order of
magnitude) in the HBB AGB star HV 2576 in the Large Magellanic Cloud 
(LMC)\footnote{Note that, to date, Al abundances have not been 
obtained in solar metallicity massive AGB stars.}. The surface 
chemistry of HV 2576 shows the imprinting of HBB, with a surface 
carbon content $\sim$10 times smaller than expected, a small depletion of 
the surface oxygen, and a +0.8 dex increase in the nitrogen abundance. 
Because aluminium is not expected to undergo any destruction process 
in AGB stars (see left panel of Fig. \ref{xal}), we interpret the strong Al 
depletion observed in HV 2576
as a further indicator for the formation of Al$_{2}$O$_{3}$ (which absorbs part 
of the gaseous aluminium available) in massive AGB stars.

\begin{figure}
\begin{minipage}{0.45\textwidth}
\resizebox{1.\hsize}{!}{\includegraphics{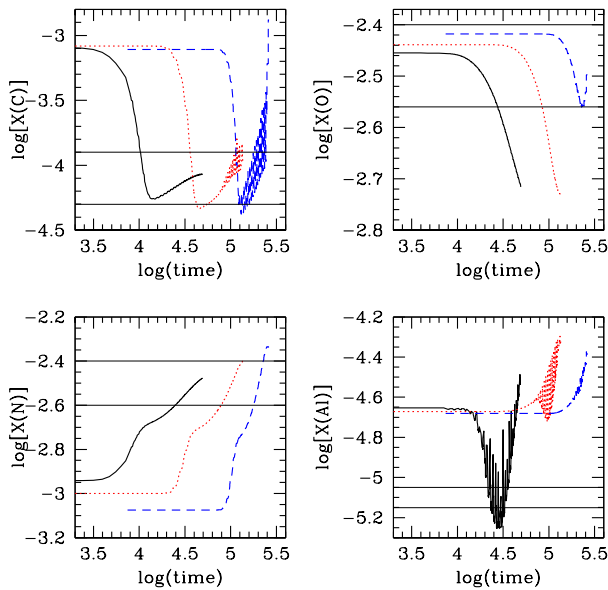}}
\end{minipage}
\vskip 20pt
\caption{The variation with time (counted from the beginning of the 
AGB phase) of the surface abundances of carbon (top-left panel), 
nitrogen (bottom-left panel), oxygen (top-right panel) and aluminium 
(bottom-right panel) of stars with initial mass of 4 (blue dashed 
line), 5 (red dotted line), and 6 M$_{\odot}$ (black solid line). The 
surface aluminium shown takes into account the amount of gaseous Al 
condensed into Al$_{2}$O$_{3}$ grains. The thin horizontal lines mark the limits 
of the abundances observed in the metal-poor HBB AGB star HV 2576 
\citep{mcsaveney}.
}
\label{2576}
\end{figure}

Figure \ref{2576} shows the variation of the surface abundances of the CNO 
elements and of aluminium in models with metallicity
Z=0.008 (appropriate for the LMC) and initial masses of 4, 5, and 6 
M$_{\odot}$  In the determination of the Al mass fraction, we subtract the 
amount of aluminium that is used to form Al$_{2}$O$_{3}$. The thin horizontal 
lines in Figure \ref{2576} indicate the upper and lower value of the abundances observed in HV 2576. 
We note the signature of HBB in the three AGB 
models, with the depletion of the surface carbon in favour of nitrogen, 
together with a small reduction of the surface oxygen.

The analysis of the predicted carbon and nitrogen abundances are of 
little help in selecting the progenitor mass (and evolutionary status) 
of HV 2576; all models achieve the CN cycle in the external envelope, 
with the consequent destruction of the surface carbon by $\sim$1 dex and 
the increase in the nitrogen content, both of them well within the 
observed range. The comparison between the observed and predicted 
oxygen abundances can be used only to rule out the possibility that HV 2576 
is a massive AGB at the latest evolutionary stage. Contrary to the CNO 
elements, the aluminium content shows a greater variation with the 
initial stellar mass and the evolutionary status on the AGB. The 
extremely low gaseous Al abundance measured in HV 2576 demands a 
considerable production of alumina dust, which is only achieved for 
the 6 M$_{\odot}$ model in the evolutionary phase of maximum Al$_{2}$O$_{3}$
production (i.e., well before the tip of the AGB). Thus, the Al 
abundance in HBB AGB stars turns out to be a good possible indicator 
of the progenitor mass and evolutionary status on the AGB.

The possible use of the Al content as a mass and/or evolutionary stage 
indicator in HBB AGB stars should be investigated in the future; e.g., 
the Al abundances could be measured in well known HBB AGB stars at 
different metallicities. Such direct comparisons of observed 
abundances and theoretical predictions will become more reliable once 
we completely understand the dust formation process and/or we know the 
Al$_{2}$O$_{3}$ sticking coefficient. However, from our comparison with HV 
2576, it is clear how the theoretical description of the AGB phase, 
and particularly of the HBB phenomenon, can potentially benefit from 
the understanding of the Al$_{2}$O$_{3}$ dust formation process in stars 
experiencing HBB.

\subsection{The Al$_{2}$O$_{3}$ dust condensation zone in Galactic O-rich AGB stars}
A valuable test for our description of the alumina dust formation 
process in O-rich AGB stars is the comparison of our results with the 
recent findings by \citet{karovicova}. These authors have 
recently analyzed a small sample of three Galactic (i.e., solar 
metallicity) O-rich AGB stars (S Ori, R Cnc, and GX Mon) by means of 
spatially and spectrally resolved mid-infrared interferometric 
observations. Two stars (S Ori and R Cnc) in their sample are 
dominated by Al$_{2}$O$_{3}$ dust and the third one (GX Mon) displays a mix of 
Al$_{2}$O$_{3}$ and Mg-silicates. Indeed, their observations indicate that the 
inner radii of the Al$_{2}$O$_{3}$ shell for the three stars is located at a 
radial distance from the center of the star of $\sim$ 2 R$_*$ (where R$_*$ is the stellar radius), 
while Mg-silicates are formed at larger radial distances. Taking into 
account all possible observational errors (e.g., the distances to 
these Galactic sources are very uncertain) as well as the 
uncertainties in the theoretical modelling (mass loss, sticking 
coefficient, etc.), these interferometric observations are in very 
good agreement (especially for Al$_{2}$O$_{3}$) with our model predictions for 
solar metallicity AGB stars (see Figure \ref{rad}).

Indeed, in our AGB wind models, the region where Al$_{2}$O$_{3}$ forms gets closer the 
stellar surface as the AGB evolution proceeds. This is due to the 
decreasing trend of the effective temperature, which is rather 
similar for models of different progenitor mass. Independently of the 
initial mass, we find that Al$_{2}$O$_{3}$ formation initially begins at 
$\sim$2.5 R$_*$, decreasing down to $\sim$1.5 R$_*$ towards the end of the AGB 
phase. Unfortunately, our predicted radii for the Al$_{2}$O$_{3}$ dust 
condensation zone cannot be used to constrain the exact mass (and 
evolutionary status) of O-rich AGB stars but the observations ($\sim$2 R$_*$) 
by \citet{karovicova} lie just between our predicted range of 
$\sim$1.5-2.5 R$_*$ for the Al$_{2}$O$_{3}$ region and seem to support the reliability 
of our dust formation description.

Furthermore, \citet{karovicova} found that the optical depth in 
the V-band for their sample stars is in the range $\tau_V \sim$ 1.3 - 2. This 
fact could rule out the possibility of these stars being AGB stars 
with initial mass above
5 M$_{\odot}$, because in our high-mass models the optical depth attains 
values $\tau_V >$ 4 during the whole thermally pulsing phase. However, 
these stars could be the descendants of stars with initial mass 
between 4 and 5 solar masses; lower mass models achieve only a modest 
(if any) production of alumina dust, whereas stars more massive than 
$\sim$ 5 M$_{\odot}$ evolve at optical depths much larger than observed by 
Karovicova et al. (2013). If the initial mass of the star is close to 
the lower limit of $\sim$ 4 M$_{\odot}$ given above, we suggest that it is 
observed in an advanced AGB stage, e.g., after $\sim$10 - 20 thermal pulses 
because $\tau_V$ is too small in the earlier AGB phases. Conversely, if 
the initial stellar mass is $\sim$ 5 M$_{\odot}$ , then we propose that the star 
is in the early AGB phase, because the optical depth
becomes too large in more advanced evolutionary stages. Indeed, based 
on some observational properties (e.g.,
the characteristics of the maser emission, variability periods, 
infrared colors, etc.; see e.g., Garcia-Hernandez et al. 2006, 2007a), 
one could argue that GX Mon is more evolved and/or more massive than S 
Ori and R Cnc.

\section{Conclusions}

We investigate the production of alumina dust (Al$_{2}$O$_{3}$) in the
circumstellar envelopes of O-rich AGB stars. We focused on those AGB stars
experiencing HBB, with masses $M\ge 3M_{\odot}$. The range of metallicities
examined is $10^{-3} \leq Z \leq 0.018$; lower metallicity stars are not
expected to produce Al$_2$O$_3$ due to the extremely low Al abundance.

Al$_{2}$O$_{3}$ is the most stable oxygen-bearing dust species and its
condensation process begins at temperatures T $\sim$ 1500 K, which is
considerably larger than that for Mg-silicates (T $\sim$ 1100 K) and iron (T
$\sim$ 1000 K). Thus, alumina dust grains form very close to the surface of
the star, at radial distances from the stellar centre ranging from $\sim$ 2
R$_{*}$ (for solar metallicity models) to $\sim$ 4 R$_{*}$ (for $Z=10^{-3}$).
This result finds a robust confirmation in the recent interferometric
observations of Galactic (i.e., solar metallicity) O-rich AGB stars by
\citet{karovicova}.

The amount of Al$_2$O$_3$ formed scales almost linearly with the metallicity,
owing to the larger surface abundances of aluminium in the higher metallicity
models. The maximum production of alumina dust occurs in massive AGBs at
solar metallicity, with a total Al$_2$O$_3$ mass of $\sim$ 10$^{-3}M_{\odot}$.
This sets an upper limit to the mass of Al$_2$O$_3$ that can be formed around
AGB stars - in the higher metallicity models the Al$_2$O$_3$ formation process
is so efficient that all the gaseous Al is absorbed and there is no
possibility for further condensation. The Al$_2$O$_3$ grain size decreases with
decreasing metallicity and progenitor mass. The maximum Al$_2$O$_3$ dust grain
size of $\sim$0.075 $\mu$m (at solar metallicity) is considered as an upper
limit to size of the alumina dust grains that can be formed around AGB stars.

The formation of alumina dust turns out to be extremely sensitive to the
initial mass of the star. Models with $M < 5 M_{\odot}$ experience soft HBB,
and thus the condensation of gaseous Al-based molecules is less efficient.
The masses of the Al$_2$O$_3$ dust formed around AGB stars of $\sim 3-4
M_{\odot}$ is $\sim 100$ times smaller than that for their more massive
counterparts. 

The amount of gaseous Al available at the stellar surface severely decreases
when alumina dust forms. Remarkably, this is consistent with the strong Al
depletion seen in the low-metallicity HBB AGB star HV 2576. We suggest that
the measurement of the Al abundances in HBB AGB stars could be potentially
used as a proxy of the strength of HBB experienced by the star (e.g., stellar
mass and evolutionary status).

Our conclusions - particularly for the predicted trend of the Al$_2$O$_3$ formed
versus the initial stellar mass - are partly affected by the specific value
of the sticking coefficient ($\alpha_{al}$=0.1, assumed to be similar to that
of Mg-silicates), which gives the efficiency of the condensation process. For
$\alpha_{al}$ values closer to unity, the Al$_2$O$_3$ formation as a function of
the progenitor mass would be much flatter because the saturation conditions
would occur even for $\sim 3-4 M_{\odot}$ AGB stars.

\section*{Acknowledgments}
The authors are indebted to the referee, Aki Takigawa, for the careful reading 
of the manuscript and for the detailed and relevant comments, that 
helped to increase the quality of this work.
D.A.G.H. acknowledges support provided by the Spanish Ministry of Economy and
Competitiveness under grants AYA$-$2011$-$27754 and AYA$-$2011$-$29060.
P.V. was supported by PRIN MIUR 2011 "The Chemical and Dynamical Evolution of the Milk Way and Local Group Galaxies" (PI: F. Matteucci), prot. 2010LY5N2T.
R.S. acknowledges funding from the European Research Council under the European Union's
Seventh Framework Programme (FP/2007-2013) / ERC Grant Agreement n. 306476.


\begin{thebibliography}{99}
\bibitem[\protect\citeauthoryear{Begemann}{1997}]{Begemann97}
Begemann B., Dorschner J., Henning Th., Mutschke H., Guertler J., Koempe C., Nass R., 1997,Apj, 476, 199
\bibitem[\protect\citeauthoryear{Bell \& Lin}{1994}]{bell94}
Bell K.R., Lin D.N.C., 1994, ApJ, 427, 987
\bibitem[\protect\citeauthoryear{Bl\"ocker}{1995}]{blocker95}
Bl\"ocker T., 1995, A\&A, 297, 727
\bibitem[\protect\citeauthoryear{Bl\"ocker \& Sch\"oenberner}{1991}]{blocker91}
Bl\"ocker T., Sch\"oenberner D., 1991, A\&A, 244, L43
\bibitem[\protect\citeauthoryear{Blommaert et al.}{2006}]{Blommaert06}
Blommaert J. A. D. L., Groenewegen M. A. T., Okumura K., Ganesh S., Omont A., Cami J., Glass I. S., Habing H. J., Schultheis M., Simon G., van Loon J. Th., 2006, A\&A, 460, 555
\bibitem[\protect\citeauthoryear{Canuto \& Mazzitelli}{1991}]{cm91}
Canuto V.M.C., Mazzitelli I., 1991, ApJ, 370, 295
\bibitem[\protect\citeauthoryear{Cherchneff \& Cau}{1999}]{cherch99}
Cherchneff, I., \& Cau, P. 1999, in {\it Proc. IAU Symp. 191, Asymptotic Giant Branch Stars}, ed. T. Le Bertre, A. Lebre, \& C. Waelkens (San Francisco, CA:ASP), p. 251 \bibitem[\protect\citeauthoryear{Choi et al.}{1998}]{Choi98}
Choi B., Huss G. R., Wasserburg G. J., Gallino R., 1998, Science, 282, 1284
\bibitem[\protect\citeauthoryear{DePew et al.}{2006}]{DePew06}	
DePew K., Speck A., Dijkstra C., 2006, Apj, 640, 971
\bibitem[\protect\citeauthoryear{Di Criscienzo et al.}{2013}]{Dic13}	
Di Criscienzo M., Dell'Agli F., Ventura P., Schneider R., Valiante R., La Franca F., Rossi C., Gallerani S., Maiolino R., 2013,
MNRAS, 433, 313 
\bibitem[\protect\citeauthoryear{Ferrarotti \& Gail}{2001}]{fg01}
Ferrarotti A.D., Gail H.P., 2001, A\&A, 371, 133
\bibitem[\protect\citeauthoryear{Ferrarotti \& Gail}{2002}]{fg02}
Ferrarotti A.D., Gail H.P., 2002, A\&A, 382, 256
\bibitem[\protect\citeauthoryear{Ferrarotti \& Gail}{2006}]{fg06}
Ferrarotti A.D., Gail H.P., 2006, A\&A, 553, 576
\bibitem[\protect\citeauthoryear{Gail \& Sedlmayr}{1985}]{gs85}
Gail H.P., Sedlmayr E., 1985, A\&A, 148, 183
\bibitem[\protect\citeauthoryear{Gail \& Sedlmayr}{1998}]{gs98}
Gail H.P., Sedlmayr E., 1998, Faraday Discussions, 109, 303
\bibitem[\protect\citeauthoryear{Gail \& Sedlmayr}{1999}]{gs99}
Gail H.P., Sedlmayr E., 1999, A\&A, 347, 594
\bibitem[\protect\citeauthoryear{Gall et al.}{2011}]{gal11}
Gall C., Hjorth J., Andersen A. C., 2011, A\&ARv, 19, 43
\bibitem[\protect\citeauthoryear{Garc\'{\i}a-Hern\'andez et al.}{2006}]{dagh06}
Garc\'{\i}a-Hern\'andez D. A., Garc\'{\i}a-Lario, P., Plez, B. et al. 2006, Science, 314, 1751  
\bibitem[\protect\citeauthoryear{Garc\'{\i}a-Hern\'andez et al.}{2007a}]{dagh07a}
Garc\'{\i}a-Hern\'andez D. A., Garc\'{\i}a-Lario, P., Plez, B. et al. 2007a, A\&A, 462, 711 
\bibitem[\protect\citeauthoryear{Garc\'{\i}a-Hern\'andez et al.}{2007}]{dagh07b}
Garc\'{\i}a-Hern\'andez D. A., Perea-Calder\'on, J. V., Bobrowsky, M., Garc\'{\i}a-Lario, P. 2007b, ApJ, 666, L33 
\bibitem[\protect\citeauthoryear{Garc\'{\i}a-Hern\'andez et al.}{2009}]{dagh09}
Garc\'{\i}a-Hern\'andez D. A., Garc\'{\i}a-Lario, P., Plez, B. et al. 2009, ApJ, 705, L31
\bibitem[\protect\citeauthoryear{Garc\'{\i}a-Hern\'andez et al.}{2013}]{dagh13}
Garc\'{\i}a-Hern\'andez, D. A., Zamora, O., Yag\"ue, A. et al. 2013, A\&A, 555, L3
\bibitem[\protect\citeauthoryear{Gomez et al.}{2012}]{gomez12}
Gomez H. L. et al., 2012, MNRAS, 420, 3557
\bibitem[\protect\citeauthoryear{Grevesse \& Sauval}{1998}]{grevesse98}
Grevesse N., Sauval A.J, 1998, SSrv, 85, 161
\bibitem[\protect\citeauthoryear{Herwig}{2005}]{herwig05}
Herwig F., 2005, AR\&A, 43, 435
\bibitem[\protect\citeauthoryear{Hutcheon}{1994}]{hutcheon94}
Hutcheon I.~D., Huss G.~R., Fahey A.~J., Wasserburg G.~J., ApJL, 425, 97
\bibitem[\protect\citeauthoryear{Jones et al.}{2014}]{Jones14}
Jones O. C., Kemper F., Srinivasan S., McDonald I., Sloan G. C., Zijlstra A. A., 2014, arXiv:1402.2485
\bibitem[\protect\citeauthoryear{Karovicova et al.}{2013}]{karovicova}
Karovicova, I., Wittkowski, M., Ohnaka, K. et al. 2013, A\&A (in press; arXiv:1310.1924)
\bibitem[\protect\citeauthoryear{Knapp}{1985}]{knapp85}
Knapp G.~R., 1985, ApJ, 293, 273
\bibitem[\protect\citeauthoryear{Koike et al.}{1995}]{Koike95}
Koike C., Kaito C., Yamamoto T., Shibai H., Kimura S., Suto H., 1995, Icarus, 114, 203
\bibitem[\protect\citeauthoryear{Levin \& Brandon}{1998}]{levin98}
Levin V., Brandon M., 1998, J.Am.Ceram.Soc., 81, 1995
\bibitem[\protect\citeauthoryear{Lugaro et al.}{2012}]{Lugaro12}
Lugaro M., Doherty C. L., Karakas A. I., Maddison S. T., Liffman K., 
Garc\'{\i}a-Hern\'andez D. A., Siess L., Lattanzio J. C., 2012, M\&PS, 47, 1998
\bibitem[\protect\citeauthoryear{Lorenz-Martins \& Pompeia}{2000}]{lore00}
Lorenz-Martins, S., \& Pompeia, L. 2000, MNRAS, 315, 856
\bibitem[\protect\citeauthoryear{McSaveney et al.}{2007}]{mcsaveney}
McSaveney, J. A., Wood, P. R., Scholz, M. et al. 2007, MNRAS, 378, 1089
\bibitem[\protect\citeauthoryear{Maldoni et al.}{2005}]{mald05}
Maldoni, M. M., Ireland, T. R., Smith, R. G., Robinson, G. 2005, MNRAS, 362, 782
\bibitem[\protect\citeauthoryear{Matsuura et al.}{2011}]{matsu11}
Matsuura M., Dwek E., Meixner M., et al. 2011, Science, 333, 1258
\bibitem[\protect\citeauthoryear{Mazzitelli et al.}{1999}]{mazz99}
Mazzitelli I., D'Antona F., \& Ventura P. 1999, A\&A, 348, 846
\bibitem[\protect\citeauthoryear{Nanni et al.}{2013}]{nanni}
Nanni A., Bressan A., Marigo P., Girardi L. 2013, MNRAS, 434, 2390
\bibitem[\protect\citeauthoryear{Nanni et al.}{2013}]{nanni}
Nittler L.~R., Alexander C.~M.~O, Gallino R., Hoppe P., Nguyen A.~N., Stadermann F.~J., Zinner E.~K., 2008, ApJ, 682, 1450
\bibitem[\protect\citeauthoryear{Norris et al.}{2012}]{nor12}
Norris, B. R. M., Tuthill P. G., Ireland M. J. et al. 2012, Nature, 484, 220
\bibitem[\protect\citeauthoryear{Posch et al.}{1999}]{Posch99}
Kerschbaum F., Mutschke H., Fabian D., Dorschner J., Hron J., 1999, A\&A, 352, 609
\bibitem[\protect\citeauthoryear{Renzini \& Voli}{1981}]{renzini81} 
Renzini A., Voli M., 1981, A\&A, 94, 175
\bibitem[\protect\citeauthoryear{Schwarzschild \& Harm}{1965}]{schw65}
Schwarzschild M., Harm R., 1965, ApJ, 142, 855
\bibitem[\protect\citeauthoryear{Schwarzschild \& Harm}{1967}]{schw67}
Schwarzschild M., Harm R., 1967, ApJ, 145, 496
\bibitem[\protect\citeauthoryear{Sedlmayr}{1989}]{sedlmayr89} 
Sedlmayr E., 1989, Interstellar Dust: Proceedings of the 135th Symposium of the 
International Astronomical Union, held in Santa Clara, California, 26-30 July 1988. 
Edited by Louis J. Allamandola and A. G. G. M. Tielens. International Astronomical Union.
\bibitem[\protect\citeauthoryear{Sharp \& Huebner}{1990}]{Sharp90} 	
Sharp C. M., Huebner W. F., 1990, Astrophysical Journal Supplement Series, 72, 417
\bibitem[\protect\citeauthoryear{Sloan}{2003}]{Sloan03} 
Sloan G. C., Kraemer Kathleen E., Goebel J. H., Price Stephan D., 2003, ApJ, 584, 493
\bibitem[\protect\citeauthoryear{Sylvester et al.}{1999}]{sylv99} 	
Sylvester, R. J., Kemper, F., Barlow, M. J. et al. 1999, A\&A, 352, 587
\bibitem[\protect\citeauthoryear{Takigawa et al.}{2009}]{takigawa09}
Takigawa A., Tachibana S., Nagahara H., Ozawa K., 2009 ,40th Lunar and Planetary Science Conference,
 held March 23-27, 2009 in The Woodlands, Texas, id.1731
\bibitem[\protect\citeauthoryear{Takigawa et al.}{2012}]{takigawa12}
Takigawa A., Tachibana S., Nagahara H., Ozawa K., 2012 ,43rd Lunar and Planetary Science Conference,
 2012, 1875
\bibitem[\protect\citeauthoryear{Takigawa et al.}{2014}]{takigawa14a}
Takigawa A., Tachibana S., Huss G. R., Nagashima K., Makide K., Krot A. N., Nagahara H., 2014, GCA, 124,309
\bibitem[\protect\citeauthoryear{Tenorio-Tagle et al.}{2013}]{ten-tag13}
Tenorio-Tagle, G., Silich, S., Mart\'{\i}nez-Gonz\'lez, S. et al. 2013, ApJ, 778, 159
\bibitem[\protect\citeauthoryear{Tielens et al.}{1998}]{tielens}
Tielens, A. G. G. M., Waters, L. B. F. M., Molster, F. J., \& Justtanont, K. 1998, Ap\&SS, 255, 415
\bibitem[\protect\citeauthoryear{Trigo-Rodr\'{\i}guez et al.}{2009}]{trigo09}
Trigo-Rodr\'{\i}guez, J. M., Garc\'{\i}a-Hern\'andez, D. A., Lugaro, M. et al. 2009, M\&PS, 44, 627
\bibitem[\protect\citeauthoryear{Valiante et al.}{2009}]{valiante09}
Valiante R., Schneider R., Bianchi S., Andersen A., Anja C., 2009, MNRAS, 397, 1661
\bibitem[\protect\citeauthoryear{Ventura, Carini \& D'Antona}{2011}]{vcd11}
Ventura P., Carini R., D'Antona F., 2011, MNRAS, 415, 3865
\bibitem[\protect\citeauthoryear{Ventura et al.}{2013}]{ventura13}
Ventura P., Di criscienzo M., Carini R., D'Antona F., 2013, MNRAS, 431, 3642
\bibitem[\protect\citeauthoryear{Ventura \& D'Antona}{2005}]{vd05}
Ventura P., D'Antona F., 2005, A\&A, 341, 279
\bibitem[\protect\citeauthoryear{Ventura \& D'Antona}{2009}]{vd09} 
Ventura P., D'Antona F., 2009, A\&A, 499, 835
\bibitem[\protect\citeauthoryear{Ventura et al.}{2012a}]{paperI} 
Ventura P., Di Criscienzo M., Schneider R., Carini R., Valiante R., D'Antona F., 
Gallerani S., Maiolino R., Tornamb\'e A., 2012a, MNRAS, 420, 1442
\bibitem[\protect\citeauthoryear{Ventura et al.}{2012b}]{paperII} 
Ventura P., Di Criscienzo M., Schneider R., Carini R., Valiante R., D'Antona F., 
Gallerani S., Maiolino R., Tornamb\'e A., 2012b, MNRAS, 424, 2345
\bibitem[\protect\citeauthoryear{Ventura et al.}{2014}]{paperIV} 
Ventura P., DellÕAgli F., Di Criscienzo M., Schneider R.,
Rossi C., La Franca F., Gallerani S., Valiante R., 2014,
arXiv:1401.1332
\bibitem[\protect\citeauthoryear{Ventura et al.}{1998}]{ventura98} Ventura P.,
Zeppieri A., Mazzitelli I., D'Antona F., 1998, A\&A, 334, 953
\bibitem[\protect\citeauthoryear{Wachter et al.}{2008}]{wachter08} 
Wachter A., Winters J. M., SchrSchršderAder K. P., Sedlmayr E., 2008, A\&A, 486, 497
\bibitem[\protect\citeauthoryear{Woitke}{2006}]{woitke06} 
Woitke P., 2006, A\&A, 460, L9
\bibitem[\protect\citeauthoryear{Yang et al.}{2004}]{yang04}
Yang X., , Chen, P., \& He, J. 2004, A\&A, 414, 1049 
\bibitem[\protect\citeauthoryear{Yang}{2008}]{yang08}
Yang X., 2008, New Astronomy, 13, 593 
\bibitem[\protect\citeauthoryear{Zhao-Geisler et al.}{2012}]{Zhao12}
Zhao-Geisler R., Quirrenbach A., K\"ohler R., Lopez B., 2012, A\&A, 545, A56
\bibitem[\protect\citeauthoryear{Zeidler et al.}{2013}]{Zeidler13}
Zeidler S., Posch Th., Mutschke H., 2013, A\&A, 553, A81
\end{thebibliography}
\end{document}